\documentclass[aoas,preprint]{imsart}

\RequirePackage{amsthm,amsmath,amsfonts,amssymb}
\RequirePackage{natbib}
\usepackage{graphicx}% uncomment this for including figures
\usepackage{bbm}
\usepackage{threeparttable}
\usepackage{lineno}
\usepackage{color}
\usepackage{xcolor}
\usepackage{soul}
\usepackage{caption}
\usepackage{subcaption}
\usepackage{lscape}
\usepackage{comment}
\usepackage{url}
\usepackage{hyperref}

\startlocaldefs
\endlocaldefs

\begin{document}

\begin{frontmatter}
%\title{On the intrinsic dimension of basketball high resolution data} 
\title{The role of intrinsic dimension in high-resolution player tracking data - Insights in basketball}

%\title{A sample article title with some additional note\thanksref{T1}}
\runtitle{On the intrinsic dimension of basketball high resolution data}
%\thankstext{T1}{A sample of additional note to the title.}

\begin{aug}
%%%%%%%%%%%%%%%%%%%%%%%%%%%%%%%%%%%%%%%%%%%%%%
%%Only one address is permitted per author. %%
%%Only division, organization and e-mail is %%
%%included in the address.                  %%
%%Additional information can be included in %%
%%the Acknowledgments section if necessary. %%
%%%%%%%%%%%%%%%%%%%%%%%%%%%%%%%%%%%%%%%%%%%%%%
\author[A,E]{\fnms{Edgar} \snm{Santos-Fernandez}\ead[label=e1,mark]{santosfe@qut.edu.au}},
\author[B,C]{\fnms{Francesco} \snm{Denti}\ead[label=e2, mark]{francesco.denti@usi.ch}},

\author[A,E]{\fnms{Kerrie} \snm{Mengersen}\ead[label=e3,mark]{k.mengersen@qut.edu.au}}
\and
\author[B,D]{\fnms{Antonietta} \snm{Mira}\ead[label=e4,mark]{antonietta.mira@usi.ch}}
%%%%%%%%%%%%%%%%%%%%%%%%%%%%%%%%%%%%%%%%%%%%%%
%% Addresses                                %%
%%%%%%%%%%%%%%%%%%%%%%%%%%%%%%%%%%%%%%%%%%%%%%
\address[A]{School of Mathematical Sciences. Y Block, Floor 8, Gardens Point Campus.
Queensland University of Technology. GPO Box 2434. Brisbane, QLD 4001. Australia. \printead{e1,e3}}

\address[E]{Australian Research Council Centre of Excellence for Mathematical and Statistical Frontiers(ACEMS)}

\address[B]{Universita della Svizzera italiana. Lugano, Switzerland, \printead{e2,e4}}
%AM I have a double affiliation: pls add University of Insubria, Italy
\address[C]{University of Milan - Bicocca. Milan, Italy} %, \printead{e2}

\address[D]{University of Insubria, Italy }
%                \printead{u3}
\end{aug}

\begin{abstract}

A new range of statistical analysis has emerged in sports after the introduction of the high-resolution player tracking technology, specifically in basketball. 
However, this high dimensional data is often challenging for statistical inference and decision making.
In this article, we employ Hidalgo, a state-of-the-art Bayesian mixture model that allows the estimation of heterogeneous intrinsic dimensions (ID) within a dataset and propose some theoretical enhancements. ID results can be interpreted as indicators of variability 
and complexity of basketball plays and games.
This technique allows classification and clustering of NBA basketball player's movement and shot charts data. 
Analyzing movement data, Hidalgo identifies key stages of offensive actions such as creating space for passing, preparation/shooting and following through.  
We found that the ID value spikes reaching a peak between 4 and 8 seconds in the offensive part of the court after which it declines.
In shot charts, we obtained groups of shots that produce substantially higher and lower successes.
Overall, game-winners tend to have a larger intrinsic dimension which is an indication of more unpredictability and unique shot placements. 
Similarly, we found higher ID values in plays when the score margin is small compared to large margin ones.
These outcomes could be exploited by coaches to obtain better offensive/defensive results. 
\end{abstract}

\begin{keyword}
\kwd{Bayesian clustering}
\kwd{intrinsic dimension}
\kwd{plays classification}
\kwd{movement data}
\kwd{shot charts}
\end{keyword}

\end{frontmatter}
%%%%%%%%%%%%%%%%%%%%%%%%%%%%%%%%%%%%%%%%%%%%%%
%% Please use \tableofcontents for articles %%
%% with 50 pages and more                   %%
%%%%%%%%%%%%%%%%%%%%%%%%%%%%%%%%%%%%%%%%%%%%%%
%\tableofcontents

%%%%%%%%%%%%%%%%%%%%%%%%%%%%%%%%%%%%%%%%%%%%%%
%%%% Main text entry area:

\section {Introduction}
\label{sec:Int} 

Basketball is a highly dynamic invasion sport, in which a team aims to score in the opposing team's basket.  
Teams use a large variety of trained plays seeking an increase in the chances of scoring. 
The introduction of the SportVU NBA player tracking technology brought player movement measurements at 25 frames per second. 
These high resolution data have motivated several spatial and spatio-temporal statistical analyses \citep[e.g.][]{goldsberry2012courtvision, Shortridge2014, cervone2016multiresolution}.    
However, such high dimensional data are often challenging for statistical inference, computationally expensive and require more sophisticated statistical techniques. 
It is well known that the placement of the players in attack and defense, and particularly the guard to the player taking the shot are related to the success of a play. 
Similarly, it is argued that teams that have more versatile players in attack produce successful shots from more unique
locations in the court i.e. they tend to have a higher players' placement variability.
Furthermore, increased movement uncertainty by the players on attack tends to be harder to defend. 
Generally, successful teams create more shooting opportunities by passing the ball more effectively. All of these factors are deemed to produce an increased success for the attacking team. 
However, effective measures of this uncertainty are yet to be developed, which could be useful to measure and improve teams and players' performance.   

Recently suggested measures like ball entropy, uncertainty and unpredictability have been regarded as key performance factors in sports games \citep{lucey2012characterizing, d2015move, skinner2015optimal, hobbs2018playing}. 
In this regard, \citet{skinner2015optimal} pointed out that the expected value in a play will decrease as it is used more often.

A large number of individual statistics are collected nowadays in basketball games.
On one hand, teams monitor the players' traditional summary statistics e.g.: the number of points (PTS), defensive rebounds (DREB), assists (AST), field goals made (FGM), 3-Point Field Goals Made (3PM), minutes (MIN), etc. 
On top of that, several other metrics are estimated from tracking technology: distance feet (Dist. Feet), average speed (Avg. Speed), passes made and received, etc.
This large number of variables times 15 players on active roster during the 82 games/season make univariate analysis and comparisons extremely laborious.

Hence, multivariate statistical techniques like clustering are becoming increasingly popular among sports scientists. 
These methods, despite their greater complexity, allow a better communication of performance to coaches. %(reduced number of dimensions )   
\citet{lutz2012cluster}, for example, used statistics such as field goals, steals and assist ratio to cluster players with similar features into 10 categories. Other clustering applications can be found in \citet{metulini2017space} and \citet{metulini2018players}.
In another example of clustering \citet{franks2015characterizing} used nonnegative matrix factorization (NMF) to group defensive players using field goal locations. This approach provides a measure of the impact of defensive players on shot frequency and probability of scoring.  

Specifically, clustering algorithms such as k-means have been used for analyzing basketball data. %movement
For instance, \citet{sampaio2015exploring}, grouped players based on performance employing attacking, defense and passing statistics.    
More recently, \citet{Nistala2019} adopted clustering for the classification of players' movement based on Euclidean distance. They clustered attacking movements into 20 groups including screen action, movement along each sideline, run along the baseline, etc.    
Other examples of applications of dimensionality reduction techniques like principal components analysis (PCA) in basketball can be found in \citet{sampaio2010effects} and \citet{teramoto2018predictive}.

Positions and movements of players on attack and defense are generally correlated since defensive players guard those in attack.
Hence, using techniques like intrinsic dimension makes possible a reduction of redundant data for statistical analysis.
However, little attention has been paid to the analysis of players' movements during possession times.  
Similarly, we found no discussion on the complexity assessment of the players' placements in shot charts and their relation to performance.  

Before delving into the analysis we add some definitions:
\begin{itemize}
    \item  A {\it possession} means to be in control of the ball.
\item The team on {\it offense}/{\it attack} refers to the team handling the ball with the aim of scoring. The team on {\it defense} is the team preventing the other one from getting points scored. 
    \item  We refer to a {\it play} as one action that starts when the team gets possession of the ball and it ends when they lose it. For example, team A gets a ball possession after team B scores. The play finishes when team A shots and misses and team B gets the rebound. The play encompasses the movements of the players during the possession. Every play has an outcome (e.g. scored, missed, etc). 
		A game is composed of a large number of plays.   
    \item  {\it Shot chart} refers here to the positions in the $x$ and $y$ axes of the 10 players when the shot was taken. This is different from the traditional shot charts that consist only of the location of the player taking the shot.  
		\item {\it Trajectory} refers to the path followed by a player during a possession. 
\end{itemize}

Specifically, the purpose of this paper is to use Bayesian statistical clustering based on the intrinsic dimension of the data to: 
\begin{enumerate}
\item assess players' movement complexities in 3-point field-goal, 
mid-range and close shots; 
\item identify the phases in the execution of a play (ball handling, creating space for passing, preparation/shooting and following through);
\item study patterns in shot charts identifying of plays that produce better outcomes; % and 
\item examine if unpredictability in attack and intrinsic dimension are linked to a better performance.  
\end{enumerate}

The rest of the article has been divided into three parts.
In the next section, we provide an introduction to the intrinsic dimension of the data and describe the data used for analysis.
This is followed by the Results section in which we examine players' trajectories followed by an analysis of shot chart data.
Finally, Section \ref{dc} concludes with a discussion of the findings and limitations.

\section{Materials and methods} 

\subsection{Intrinsic dimension of the data}

The ID of a dataset can be defined as the dimension $d$ of the latent manifold in which the statistical units, observed in a $D$-dimensional space, lie. Generally, we expect some degree of dependency among the variables of a dataset, therefore usually $d<D$.
More formally, we refer to the definition of ID provided by \citet{Bishop95}: ``a set in $D$ dimensions has an intrinsic dimension equal to $d$ if the data lies within a $d$-dimensional subspace of $\mathbb{R}^{D}$ entirely, without information loss.''
Estimating the ID of a dataset is crucial for subsequent dimensionality reduction analyses.
In practice, high-dimensional datasets can be projected onto a subspace of smaller dimension without losing much information \citep{levina2005maximum, camastra2016intrinsic}. 
Therefore, the ID gives an indication of the complexity, redundancy and unique features in a dataset. 

Several methods for estimating the ID of data have been developed. The literature in this field is vast, but we refer to \citet{Campadelli2015} for a comprehensive review. Recently, \citet{facco2017estimating} suggested a local model-based ID estimator, exploiting results from Poisson Processes theory. The approach relies on the following results. Consider a dataset of $N$ observations $\mathbf{X}=\{\mathbf{x}_i\}_{i=1}^N$, with $\mathbf{x}_i \in \mathbb{R}^D$. Moreover, let $\mathbf{x}_{(j,i)}$ be the $j$-th nearest neighbor (NN) of the $i$-th observation. Consider a distance function $d:\mathbb{R}^D\times\mathbb{R}^D\rightarrow\mathbb{R}^+$ (e.g. the Euclidean distance) and let $r_{ij}=d(\mathbf{x}_{i},\mathbf{x}_{(j,i)})$ be the distance between the $i$-th observation and its $j$-th NN. Then, it can be proven that, if the density of the data points is assumed to be constant on the scale of the second NN,

\begin{equation}
    \mu_i = \frac{r_{i2}}{r_{i1}} \sim Pareto(1,d),
\label{eq::firstresult}    
\end{equation}
where $d$ is the ID of the data. Recall that a Pareto random variable $X$ with scale parameter $a$ and shape parameter $d$ is characterized by a density of the following form: $f_X(x)=\frac{d a^{d}}{x^{d+1}}$.
\citet{facco2017estimating} proposed a classical estimator based on a linearized least squared estimator. %FD removed maximum likelihood approach.
\citet{allegra2019clustering} further extend this method, developing a Bayesian mixture model of $K$ Pareto distributions:
\begin{equation}
    \begin{aligned}
  P\left ( \mu_i | \mathbf{d},\mathbf{p} \right ) %\doteq \sum_{k=1}^{K} p_k \textrm{Pareto}\left(d\right) 
  &\doteq \mathcal{P}(\mu_i)=\sum_{k=1}^{K}  p_k d_k \mu_{i}^{-\left(d_k+1\right)}  \\
  \mathbf{p}=\left(p_1\ldots, p_K\right) &\sim Dir(c_1,\ldots,c_K),\\  d_k &\sim Gamma(a,b), \:\:\: \forall k=1,\ldots K,
    \end{aligned}
    \label{eq::model}
\end{equation}
where the vector of mixture weights is modeled with a Dirichlet distribution and the components parameters, i.e. the different IDs, are chosen to be Gamma distributed. Both 
prior choices are motivated by conjugacy, that greatly simplifies posterior simulation and inference.
Notice that, in this manner, we are able to detect heterogeneous 
IDs in a dataset characterised by multiple hidden manifolds.\\ 
A set of auxiliary variables, indicating the cluster membership for each observation (i.e. the manifold assignment), $\mathbf{z}=\left(z_1,\ldots,z_N\right)$ is introduced. Consequently, the first line of \eqref{eq::model} is rewritten as:
\begin{equation}
\mu_i|z,\mathbf{d} \sim Pareto(1,d_{z_i}), \quad z_i|\mathbf{p} \sim \sum_{k=1}^K\pi_k\delta_k(z_i), %Categ(p_1,\ldots,p_K).
\end{equation}
where $\delta_x(y)$ is the usual Dirac delta, equal to 1 if $x=y$, and 0 otherwise.

The clustering induced in the data by the latent variable $\mathbf{z}$ plays a key role: the observations within each group concur to the estimation of a different value of $d_{z_i}$. The estimation of the manifold assignment is a delicate task. Hence, if the clustering is inaccurate, so is the estimate.
Unfortunately, the Pareto densities overlap to a great extent, even for very different values of the shape parameters $\mathbf{d}$: multiple Paretos can be a viable choice for the same data point. This implies that the correctness of the cluster membership is jeopardized. 
To solve this issue, the authors consider the following additional assumption: \emph{the different manifolds are separated in the space, and the neighborhood of a point should be more likely to contain points sampled from the same manifold than points sampled from a different manifold}. To reflect this in the statistical model, they add an extra term in the likelihood, modeling the adjacency structure among the data aiming at enforcing local homogeneity.\\ 

In detail, the authors introduce the $N\times N$ adjacency matrix $\mathcal{N}^{(q)}$. The $(i,j)$ entry $\mathcal{N}_{i j}^{(q)}$ of this binary matrix is equal to one only if the observation $j$ is one of the first $q$ NNs of observation $i$, zero otherwise. Notice that $\sum_{j} \mathcal{N}_{i j}^{(q)}=q$. To induce local uniformity, 
they model $f\left(\mathcal{N}_{i j}^{(q)}=1 | z_{i}=z_{j}\right) \propto \zeta_0,$ and 
$f\left(\mathcal{N}_{i j}^{(q)}=1 | z_{i}\neq z_{j}\right) \propto \zeta_1$, where the probabilities $\zeta_0,\zeta_1$ are such that $\zeta_0 > 0.5$ and $\zeta_1<0.5$. These inequalities imply that points assigned to the same manifold have more chances to be neighbors. For simplicity, we set $\zeta_0=\zeta$ and $\zeta_1=1-\zeta$.
Denote with $\mathcal{N}_i^{(q)}$ the $i$-th row of the adjacency matrix. If we regard the rows of $\mathcal{N}^{(q)}$ as independent, the new likelihood can be written as
\begin{equation}
f\left(\mathcal{N}^{(q)} | \mathbf{z}, \zeta\right)=\prod_{i} f\left(\mathcal{N}_{i}^{(q)} | \mathbf{z}, \zeta\right)=\prod_{i} \frac{\zeta^{n_{i}^{in}(\mathbf{z})}(1-\zeta)^{q -n_{i}^{i n}(\mathbf{z})}}{\mathcal{Z}\left(\zeta, N_{z_i}\right)},
    \label{distrN}
\end{equation}
where $\zeta\in \left(0.5,1\right)$ is the parameter enforcing uniformity between neighbors ($\zeta=0.5$ implies no additional term in the likelihood), $n_{i}^{i n}(\mathbf{z})=\sum_{j} n_{i j} \mathbb{I}_{z_{j}=z_{i}}$ is the number of the $q$ NNs of $\mathbf{x}_i$ that are clustered together with observation $i$, $N_{z_i}$ is the cardinality of cluster of instances grouped with $\mathbf{x}_i$ and $\mathcal{Z}\left(\zeta, N_{z_i}\right)$ is the normalizing constant. 
The resulting likelihood for $\boldsymbol{\mu}=\left(\mu_1,\ldots,\mu_n\right)$ is 
\begin{equation}
    \mathcal{L}\left(\boldsymbol{\mu}|\mathbf{d},\mathbf{z},\zeta\right) = \prod_{i=1}^{n} \mathcal{P}\left(\mu_i|d_{z_i}\right)\times f\left(\mathcal{N}_{i}^{(q)} | \mathbf{z}, \zeta\right).
\label{MODpara}
\end{equation}

The number of mixture components is chosen ex-post, adopting some postprocessing procedure. \citet{allegra2019clustering} compare the average log-posterior estimated over the MCMC sweeps. %FD removed: Another approach could be comparing the average posterior values for different $K$. 
Alternatively, one could use more complete measures of model comparisons, like DIC, BIC, AICm, BICm, or WAIC.
The posterior distribution for the parameters cannot be obtained analytically but is approximated by Markov chain Monte Carlo simulations.

\subsection{Model Enhancements}
In this article, we suggest two types of enhancements to the work of \citet{allegra2019clustering} that help to recover more meaningful posterior inference. More precisely, our contribution is twofold. First, we propose to modify the prior on $\mathbf{d}$. Secondly, we suggest to exploit the richness of the MCMC output in a more complete way, recovering an estimate of the ID for each observation. Such an estimate will turn out to be extremely useful to interpret results in the analysed framework. Let us focus on different prior specification for the ID first.\\

\textbf{Truncated Prior}. It can happen that, especially when $D$ is low, part of the posterior distribution of $d_{z_i}$ falls above the maximum dimension $D$. To correct this issue, we propose to substitute the Gamma prior on $d_k$ with a Truncated Gamma over $(0,D)$, i.e.

$$\pi(d_k) \propto \frac{b^{a}}{\Gamma(a)} d_k^{a-1} \exp\{-b d_k\} \mathbbm{1}_{(0,D)} \quad \forall k.$$

Alternatively, if we want to include the case where $d=D$, we can employ the following density, with mixture proportion $\hat\rho$:
$$\pi(d_k) \propto \hat{\rho} \frac{b^{a}}{\Gamma(a)} d_k^{a-1} \exp\{-b d_k\} \mathbbm{1}_{(0,D)} + (1-\hat\rho) \delta_D(d_k)  \quad \forall k.$$

\textbf{Repulsive Prior}. Dealing with mixture models could lead to overfitting, in the sense that the model tends to create more components than the ones that are actually needed. Then, in some applications one may observe different clusters of observations characterized by very similar IDs. This distinction, instead of reflecting a real difference in the latent manifold dimensions, could be simply due to noise in the observed data or small curvatures in the latent geometry.
To avoid the creations of redundant components and shrink to zero the fluctuations in the estimation, we employ a repulsive density of the following form as in \citet{petralia2012repulsive}: %Petralia et al. (2012):
\begin{equation}
    \begin{aligned}
\pi(\mathbf{d}) &= c_{1}\left(\prod_{k=1}^{K} g_{0}\left(d_{k}\right)\right) h(\mathbf{d}), \quad \quad
h(\mathbf{d})=\min_{\{(s, j) \in A\}} g\left(\Delta\left(d_{s}, d_{j}\right)\right)
    \end{aligned}
\end{equation}

where $\Delta$ is a suitable distance in $\mathbb{R}^+$, $g_0$ is a univariate density function for $d_k$, and $A=\{(s,j):s=1,\ldots,K; j<s\}$. Instead of specifying the function $g$ as in the aforementioned paper, i.e. $g\left(\Delta\right)=\exp \left[-\tau\left(\Delta\right)^{-\nu}\right]$ with $\tau,\nu>0$, we adopt the following sigmoidal function:

\begin{equation}
g(\Delta)=\frac{1}{1+\exp\left[-\frac{\Delta-\tau}{\nu}\right]} \quad \tau,\nu>0.
\end{equation}

\begin{figure}[ht]
    \centering
    \includegraphics[scale=.5]{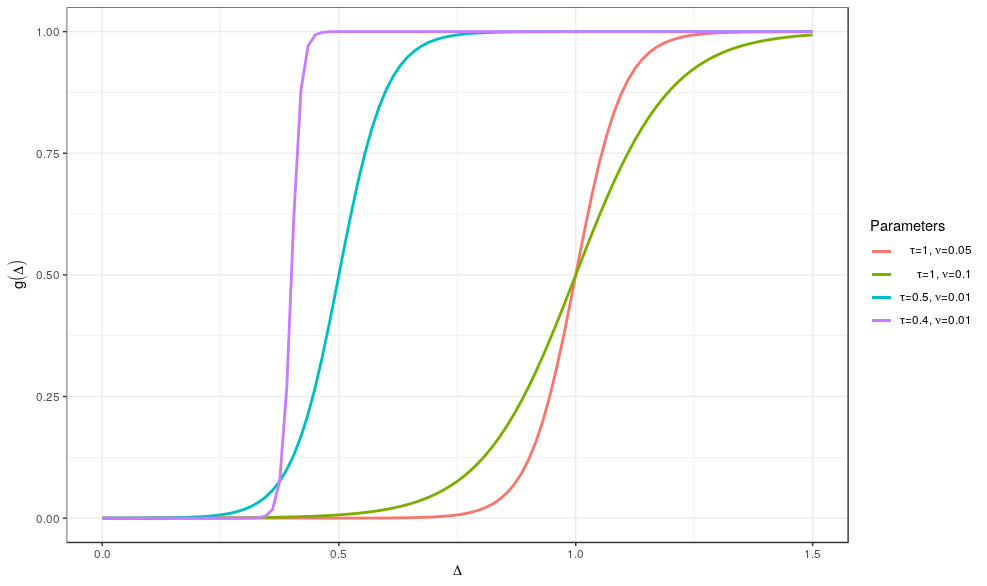}
    \caption{Different functions $g(\Delta)$ for different configurations of parameters $\tau$ and $\nu$.}
    \label{fig:gDelta}
\end{figure}

This sigmoidal function is convenient because it allows to directly specify the magnitude of the repulsion. For $\nu\rightarrow 0$ the sigmoid approaches a step function, where the jump is exactly at $\tau$. In other words, choosing a parameter $\nu$ small enough, we can induce a distance of at least $\tau$ between the realizations of the vector $\mathbf{d}$. Figure \ref{fig:gDelta} reports some examples of the function $g(\Delta)$ for different choices of $\tau$ and 
$\nu$.

\subsection{Posterior Inference}

The main goal of this methodology is to recover meaningful clusters of observations and estimate the IDs therein. We derive the ID on each observation using the following estimators, where $T$ denotes the number of MCMC sweeps:
\begin{equation}
\hat{d}_i=\frac{1}{T}\sum_{t=1}^{T} d_{z^t_{i}},\quad \quad \hat{d}_i=\text{median} \{d_{z^t_{i}}\}_{t=1}^T.
\label{mean::median}
\end{equation}
Tracking the chains of parameters actually assigned to each observation via $z^t_i$ is a simple way to deal with the label switching problem. We are well aware that more complete and involved methodologies have been proposed for handling this non-identifiability issue \citep{Robert2010,Rodriguez2014a,Celeux1998,Sperrin2010,Fruhwirth-Schnatter2011} and we leave the adoption of refinements for future research. Then, let us focus on how to recover meaningful partition. We recover the optimal partition minimizing suitable loss functions, such as the Binder loss or the Variation of Information \citep{Binder1978,Wade2015}, computed on the posterior pairwise coclustering matrix between observations. However, the model-based clustering recovered in this way may suffer from the overlapping among the Paretos in the likelihood and consequently 
might not be reliable. Another simple solution is to derive a clustering structure by inspecting the MCMC posterior median estimates as in Eq.\ref{mean::median}. To estimate an interesting partition, we can apply classical clustering algorithms such as $k$-means, where the optimal number of groups can be fixed studying the behavior of cluster quality indexes such as \emph{Silhouette} \citep{Rosseeuw1987} or the \emph{Calinski-Harabasz index} \citep{CalinskiT.1974}.

\subsection{ Description of the dataset} 
We used STATS SportVU high-resolution player tracking raw data from the NBA during the season 2015-16. We obtained the play-by-play events description and other statistics from the official website \url{https://stats.nba.com/}. %
The match between event in these two files was verified via manual video annotation of the game available on \url{https://www.youtube.com/}. 
The raw movement data needed curation and manual matching which is time-consuming. Therefore for the purpose of this research, we considered 15 random games (Table~\ref{table:games}).  

We carried out several analyses assessing the ID:

\begin{enumerate}
    \item within plays: It is based on the movement of the players during a ball possession. Section \ref{sec:md}. 
    \item between plays: using shot chart data composed by the locations of the players at the moment of the shot. Section \ref{sec:charts}.
    \item between games, using also shot chart data as in 2. Section \ref{sec:charts}
\end{enumerate}

\begin{table}[hptb]
\centering
\caption{15 randomly selected games from season 2015-16}
\begin{tabular}{llll}
\hline
Away & Home & Date(MM.DD.YYYY) & Result  \\\hline
GSW  & LAL  & 01.05.2016       & 109-88  \\
MIL  & CHI  & 01.05.2017       & 106-117 \\
MIA  & TOR  & 01.22.2016       & 81-101  \\
CLE  & GSW  & 12.25.2015       & 83-89   \\
HOU  & SAS  & 01.02.2016       & 103-121 \\
PHI  & LAL  & 01.01.2016       & 84-93   \\
MEM  & OKC  & 01.06.2016       & 94-112  \\
UTA  & SAS  & 01.06.2016       & 98-123  \\
BKN  & BOS  & 01.02.2016      & 100-97 \\
%BOS  & NYK  & 01.12.2016       & 114-120 \\
TOR  & CLE  & 01.04.2016       & 100-122 \\
MIA  & GSW  & 01.11.2016       & 103-111 \\
OKC  & CHA  & 01.02.2016       & 109-90  \\
MIA  & WAS  & 01.03.2016       & 97-75   \\
MIA  & PHX  & 01.08.2016       & 103-95  \\
GSW  & POR  & 01.08.2016       & 128-108\\\hline

\label{table:games}
\end{tabular}
\end{table}

\begin{figure}[ht]
\centering
\begin{subfigure}{.5\textwidth}
  \centering
	\caption{Video frame at the moment of the shooting.}
  \includegraphics[width=1.2\linewidth]{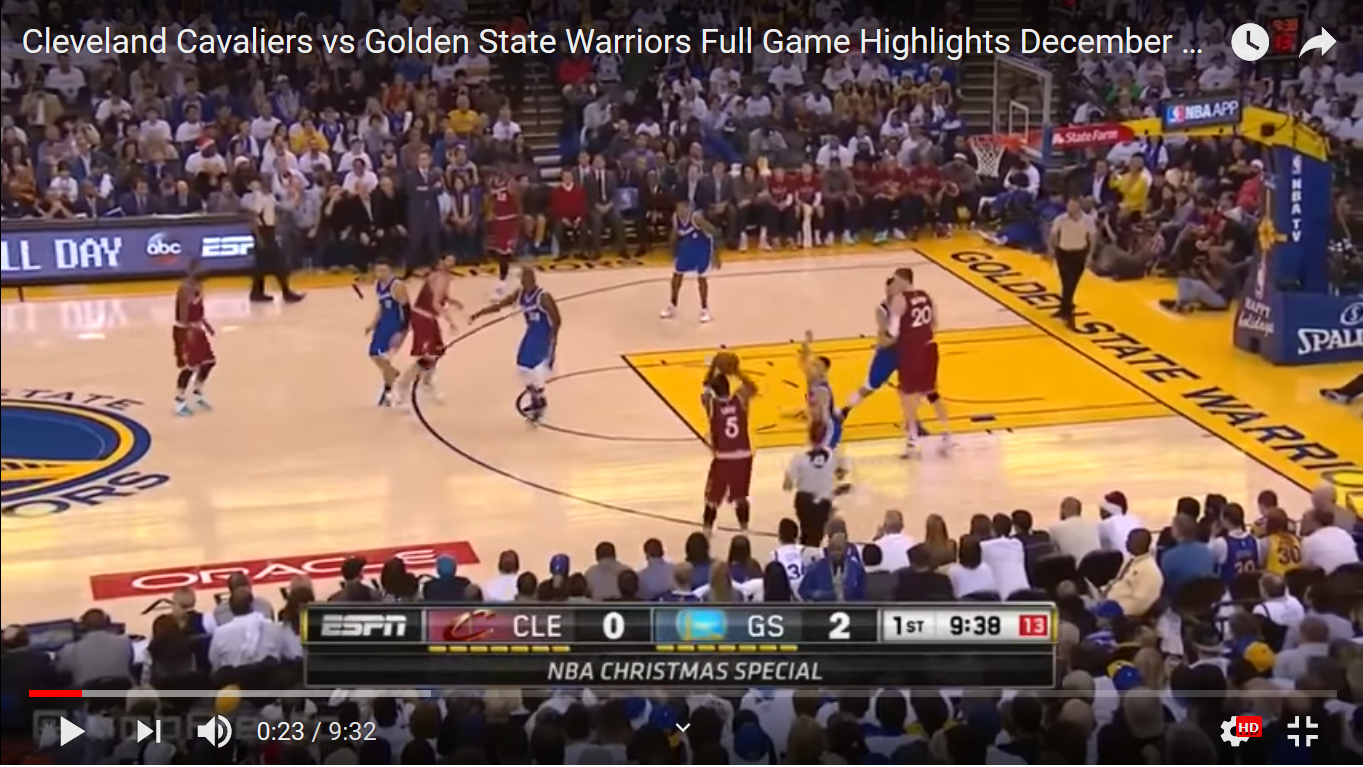}
  \label{fig:sub1}
\end{subfigure}%
\begin{subfigure}{.5\textwidth}
  \centering
\caption{Locations from the high resolution player tracking technology.} 
 \includegraphics[width=1.12\linewidth]{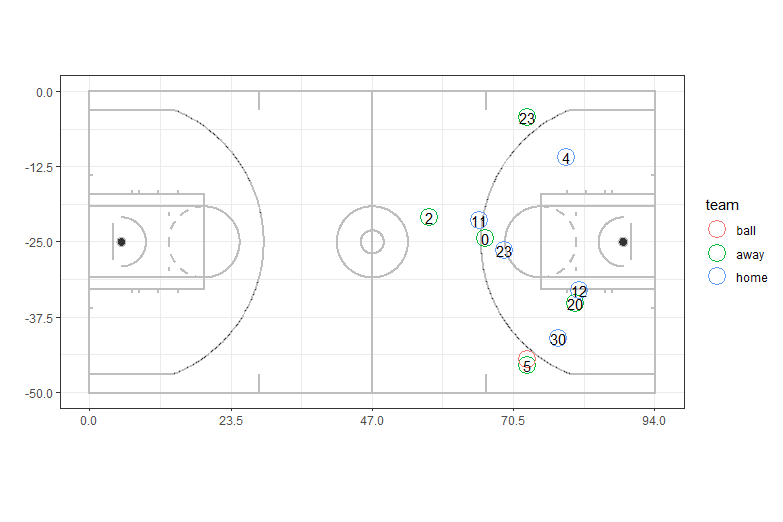}
  \label{fig:sub2}
\end{subfigure}
\caption{Locations of the players and the ball for the first scored three-point field goal of the game Cleveland Cavaliers (CLE) and the Golden State Warriors (GSW) on the $25^{th}$ of December, 2015.  This play can be watched at \url{https://youtu.be/jb57MFQLoRo?t=17}}
\label{fig:bothfig}
\end{figure}

We illustrate the application of the ID method using the game between Cleveland Cavaliers (CLE) and the Golden State Warriors (GSW) on the $25^{th}$ of December 2015. These teams made it to the final in that season. 
Fig. \ref{fig:bothfig} shows the locations of the players during the first scored three-point field goal of the game by the video screen-shot (a) and the representation of play obtained from the high-resolution player tracking technology (b).

\clearpage

\section{Results}

\subsection{ID in the Analysis of Movement Data}
\label{sec:md}
In this section, we assess the change in ID within plays produced by players' movement data in the offensive court i.e. after the ball passes the 47-foot central line, which is a current practice e.g. \citet{franks2015characterizing}.

The resolution of each play is reduced from 25 frames/second to 2.5 for faster computation without losing a substantial amount information. 
Frame 1 corresponds to the first timestamp during the play and the number of frames in a play is based on the duration. See the animation of the play from Fig. \ref{fig:bothfig} in \url{https://github.com/EdgarSantos-Fernandez/id_basketball}.
Data and the R codes for the ID computation can be found also in this repository. 

\begin{figure}[ht]
    \centering
    \includegraphics[scale=.75]{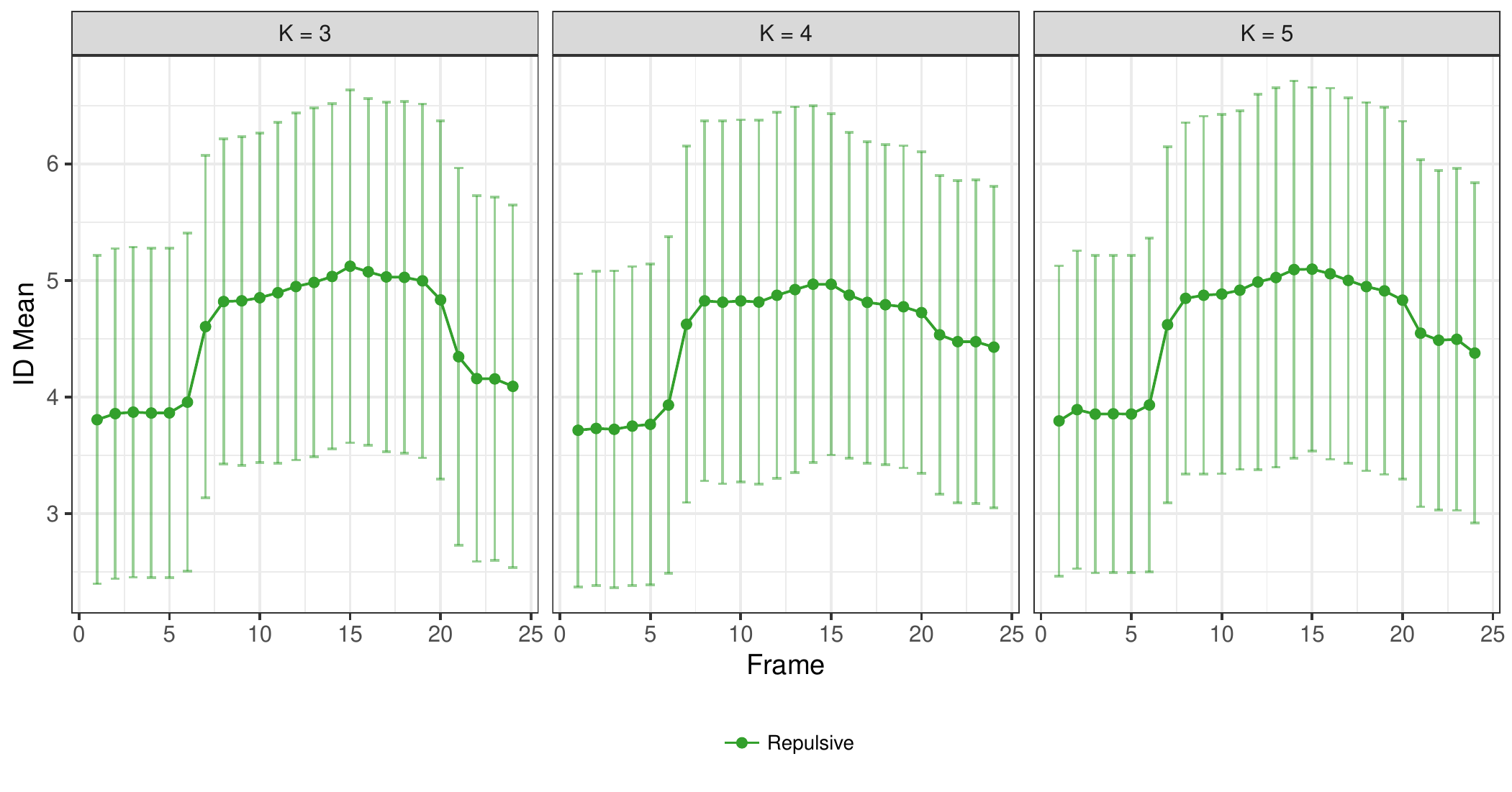}
    \caption{Comparison of the posterior ID median values plus error bars on the movement data from the play in Fig. \ref{fig:bothfig} using several mixture components.}
    \label{fig:K_method_ID_repulsive}
\end{figure}

A simulation study was carried out to assess the impact of the three priors on the posterior ID values. We found consistent results in the posterior ID median regardless of the prior as we show in Fig. \ref{fig:K_method_ID}.
Similarly, the results were stable independently of the number of chosen components ($K$). For example, Fig. \ref{fig:K_method_ID_repulsive} illustrates the evolution of the ID within a play for $K = 3,4 $ and 5. From this point on we use the repulsive variant in our analysis with $K = 3$.

\begin{figure}[htp]
\centering
\begin{subfigure}[b]{0.5\textwidth}
  \centering
	\caption{Movement of three players and the ball during a play.} 
		\includegraphics[width=3.15in]{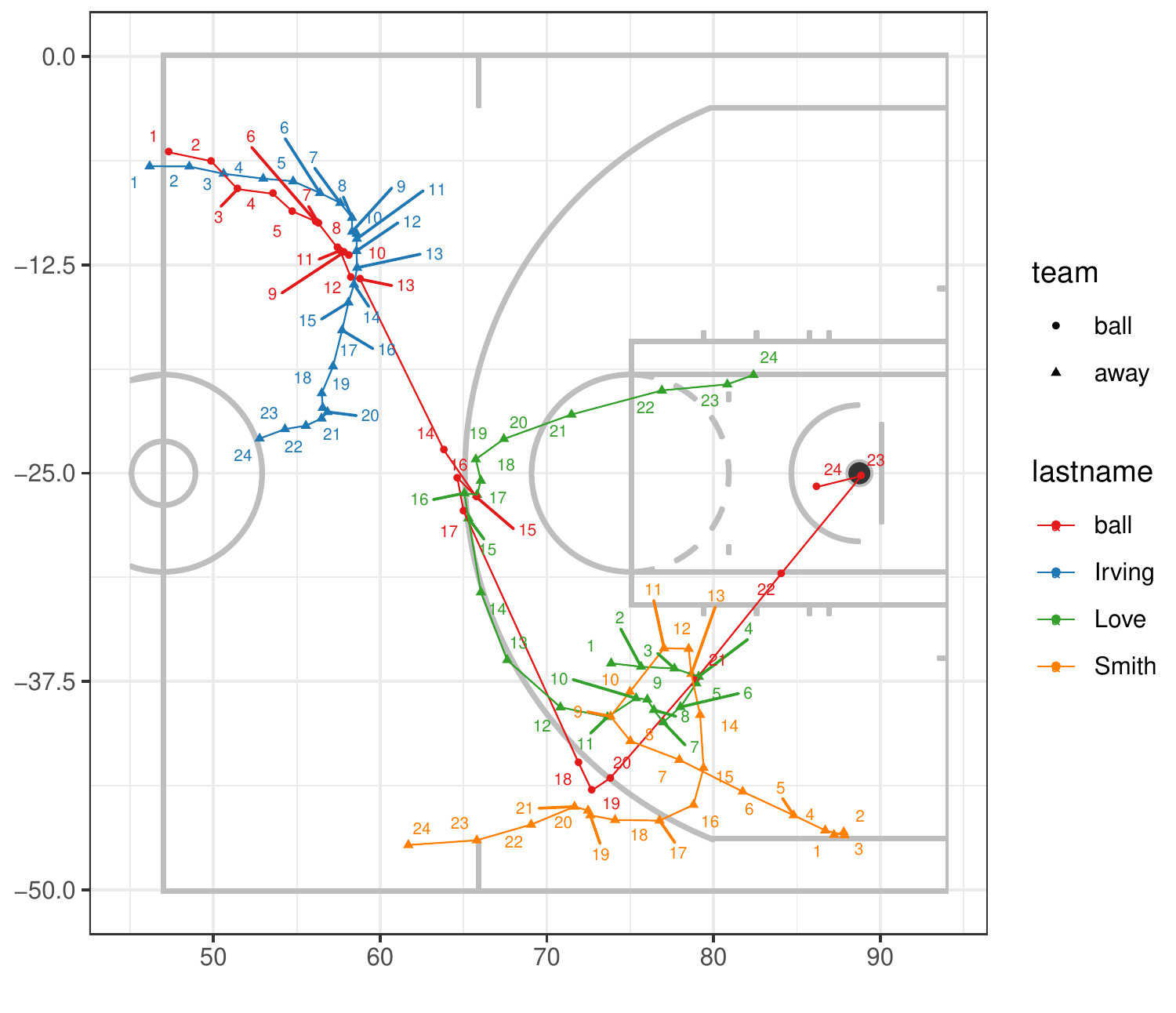}
  \label{fig:id15traj}
\end{subfigure}%
\begin{subfigure}[b]{0.5\textwidth}
  \centering
	\caption{Heatmap of the 10 players movements during the 24 time stamps and posterior medians of the ID.} 
	\vspace{-0.85cm}
		\includegraphics[width=3.15in]{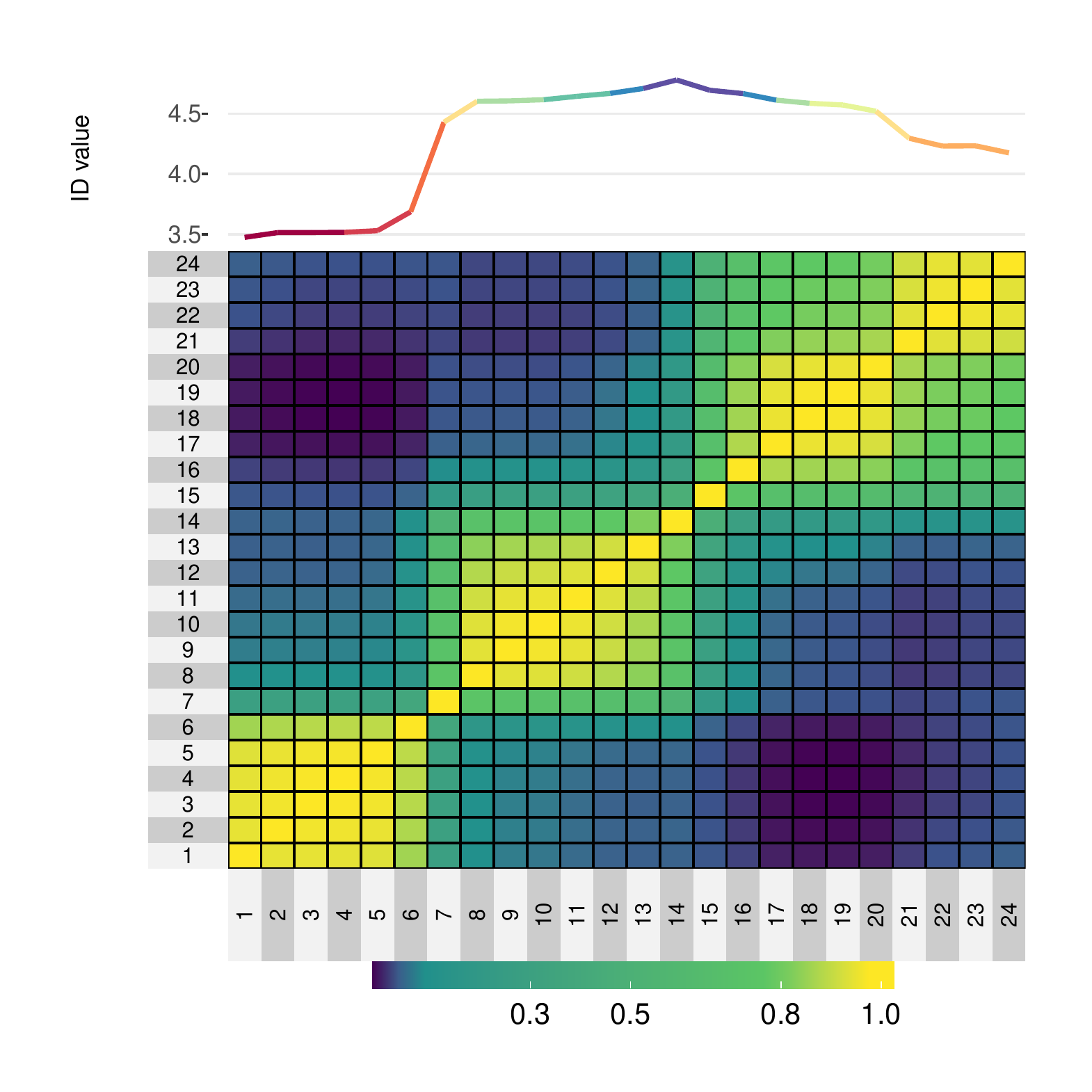}
  \label{fig:id15ID}
\end{subfigure}
\caption{Trajectory of the three attacking players involved in the play plus the ball, heatmap of the coclustering matrix and ID for the first scored three-point field goal of the game by CLE. The frequency was set at 2.5 frames/second and the play is composed of 24 frames. \url{https://youtu.be/jb57MFQLoRo?t=17}}
\label{fig:both}
\end{figure}

We analyzed all the plays during the GSW vs CLE game mentioned above computing the posterior ID distributions.
In Fig. \ref{fig:id15traj} we show the trajectory of the 3 players involved in the play illustrated in Fig. \ref{fig:bothfig}.
In this play, Irving crosses the center line dribbling and in frame 7 the players start seeking space receiving the pass. 
In frame 13 the ball goes to Love who passed to Smith and this one executed a three-pointer in frame 19. 
Fig. \ref{fig:id15ID} gives the heatmap of the posterior similarity matrix obtained from the Bayesian ID estimation algorithm using the locations ($x,y$) of the 10 players in the court. 
The line plot on top of Fig. \ref{fig:id15ID} represents the progression of the median ID across the 24 frames of play.
The evolution in ID captures changes in movement dynamics and complexities within a play. 
We note a spike in the ID value produced in stamp 7.
As expected consecutive frames are highly clustered since players tend to preserve the momentum in short intervals of times, but some interesting changes can be observed as in frame 7 and 15. 
With this approach we can identify the stages: ball handler crossing the center line (frames: 1-6), creating space for passing (frames: 7-12), preparation/shooting (frames: 17-20) and following through (frames: 21-24). 
Another example of movement analysis during a \emph{driving bank} 2-points shot by Kyrie Irving is presented in Appendix \ref{sec:ap21}.

\begin{figure*}[thp]
	\centering
			\includegraphics[width=3.5in]{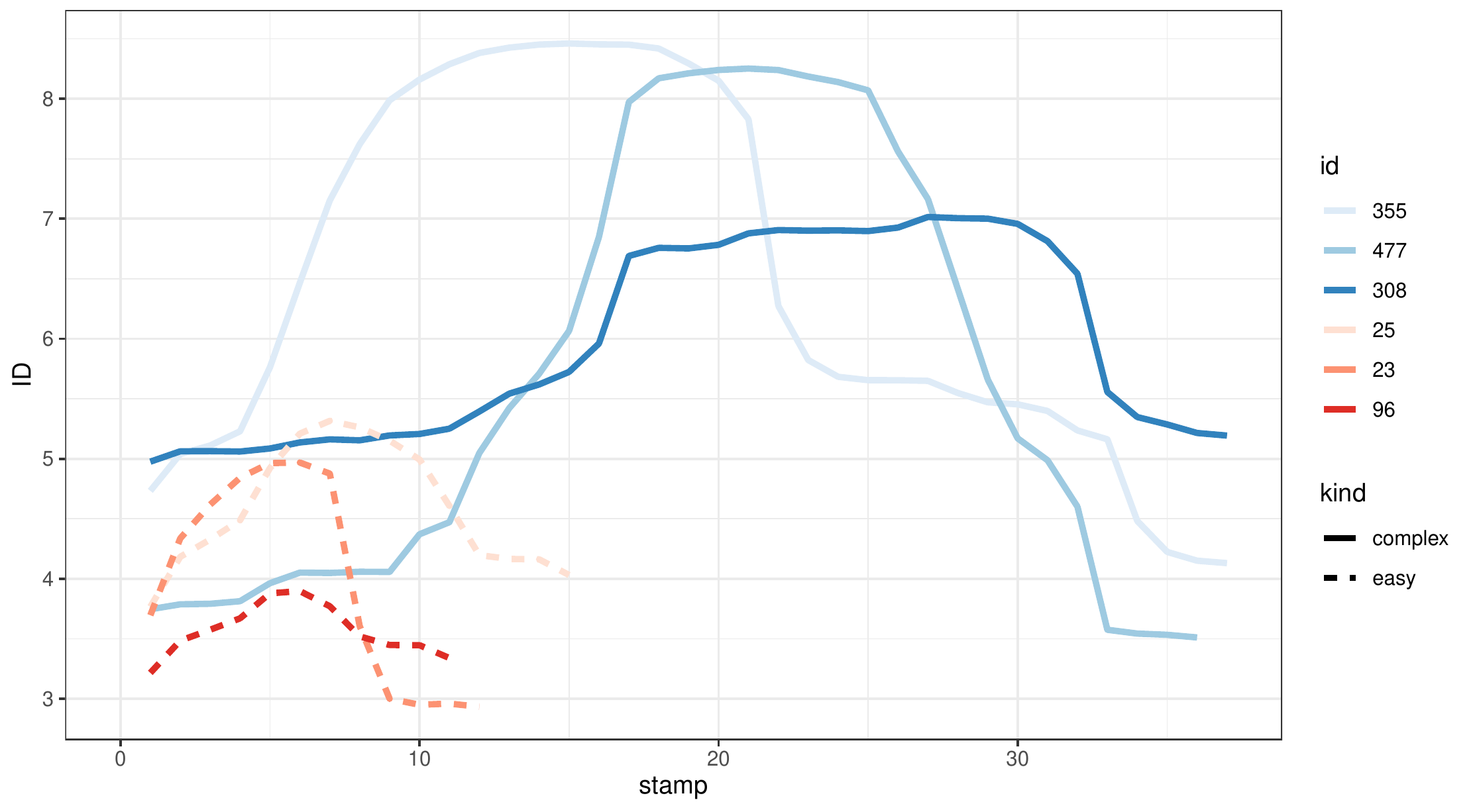}
	\caption{Example of the ID in six different plays, three complex (solid line) and three simple plays (dotted line).	} 
	\label{fig:ComplexSimplePlays}
\end{figure*}

\begin{figure}[htp]
\centering
\begin{subfigure}{.5\textwidth}
  \centering
	\caption{Simple play (\url{https://www.youtube.com/watch?v=jb57MFQLoRo&t=91s}).} 
	\vspace{-0.6cm}
		\includegraphics[width=3.25in]{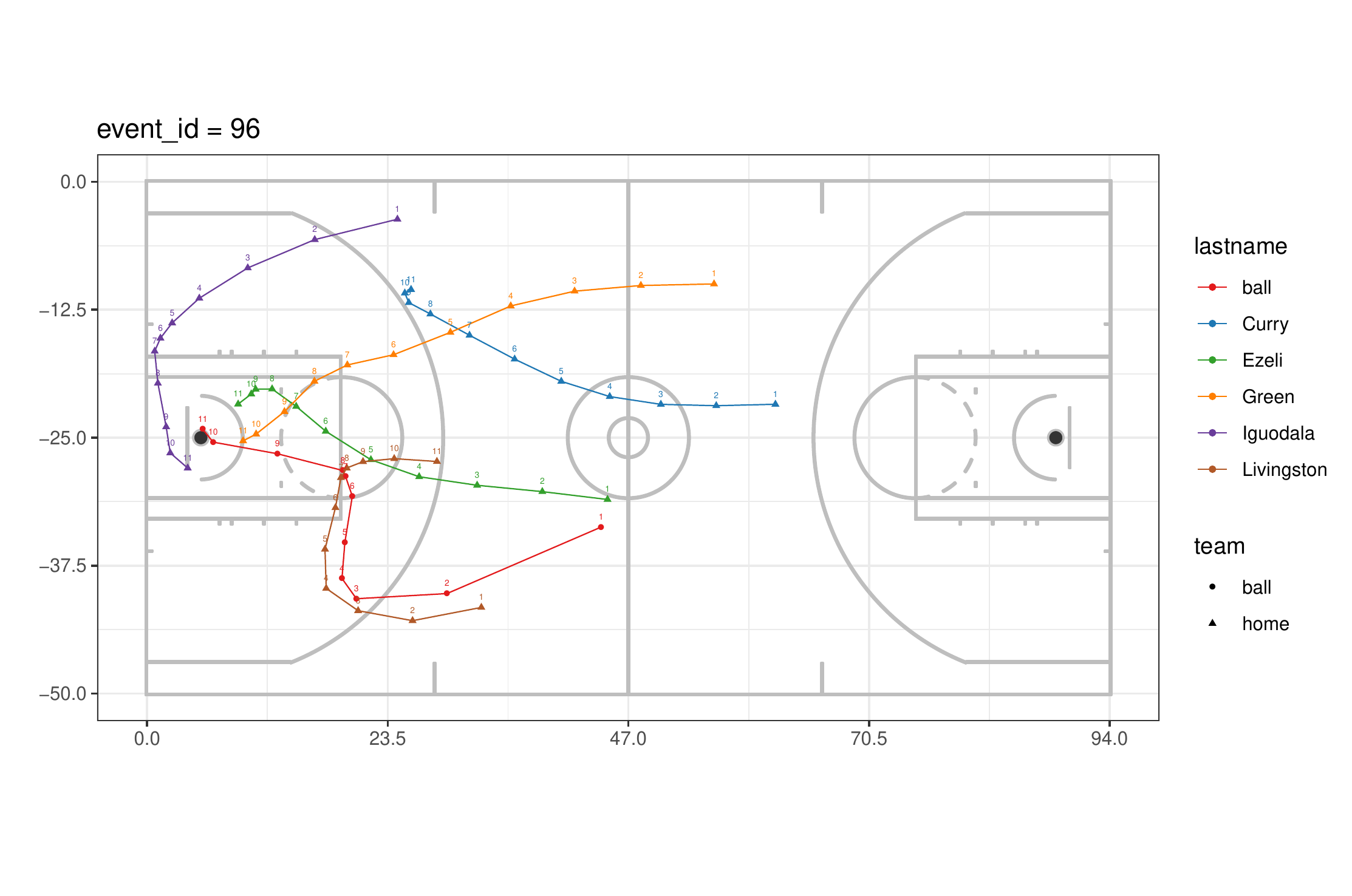} %id25
		\label{fig:simple}
\end{subfigure}%
\begin{subfigure}{.5\textwidth}
  \centering
	\caption{Complex play (\url{https://www.youtube.com/watch?v=jb57MFQLoRo&t=227s}).} 
	\vspace{-0.6cm}
\includegraphics[width=3.25in]{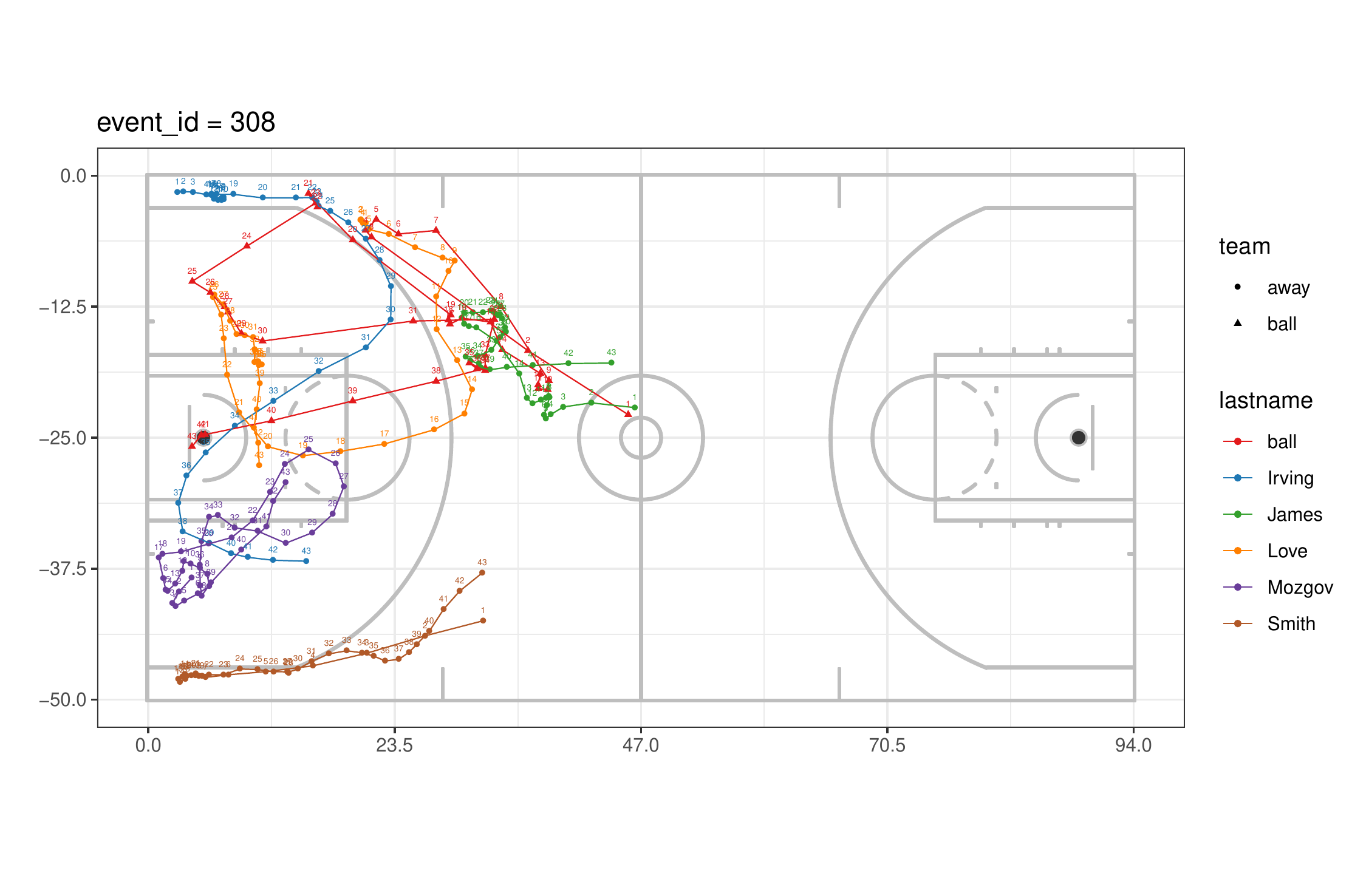} %id145
		\label{fig:complex}
\end{subfigure}
\vspace{-1cm}
\caption{Example of players trajectories (indexed by \texttt{last name}) in simple and complex plays.}
\label{fig:bothfig2}
\end{figure}

Possessions, where players move in the same direction, tend to have a smaller median ID value, while higher values are generally obtained in multi-directional trajectories and complex plays. 
We illustrate this principle in the following example. 
We consider three simple plays ($idn$ = 23, 25, 96) and three complex ($idn$ = 355, 477, 308), where $idn$ is the play identification number. 
The median ID value across these plays is given in Fig.~\ref{fig:ComplexSimplePlays}.
Simple plays are in general shorter where the players reach easily the painted zone without much resistance.
Fig.~\ref{fig:simple} and \ref{fig:complex} show an example the trajectories of the five players on attack plus the ball in plays 96 and 308.
A link to the video is also provided.

%%%%%%%%%%%%%%%%%
We performed a further analysis based on the Euclidean distance ($\delta$) from the player taking the shot to the hub. Three shot groups were defined as follows:
\begin{itemize}
    \item short distance shots (dunks, tips, etc.). Where $\delta < 6$ feet.
    \item mid-range shots (short and long two-points shots). Where $6 \leqslant \delta < 22$ feet.
    \item 3-points shots (shots from behind the line). Where $\delta \geqslant 22$ feet.
\end{itemize}

\begin{figure*}[!tph]
	\centering
		\includegraphics[width=4in]{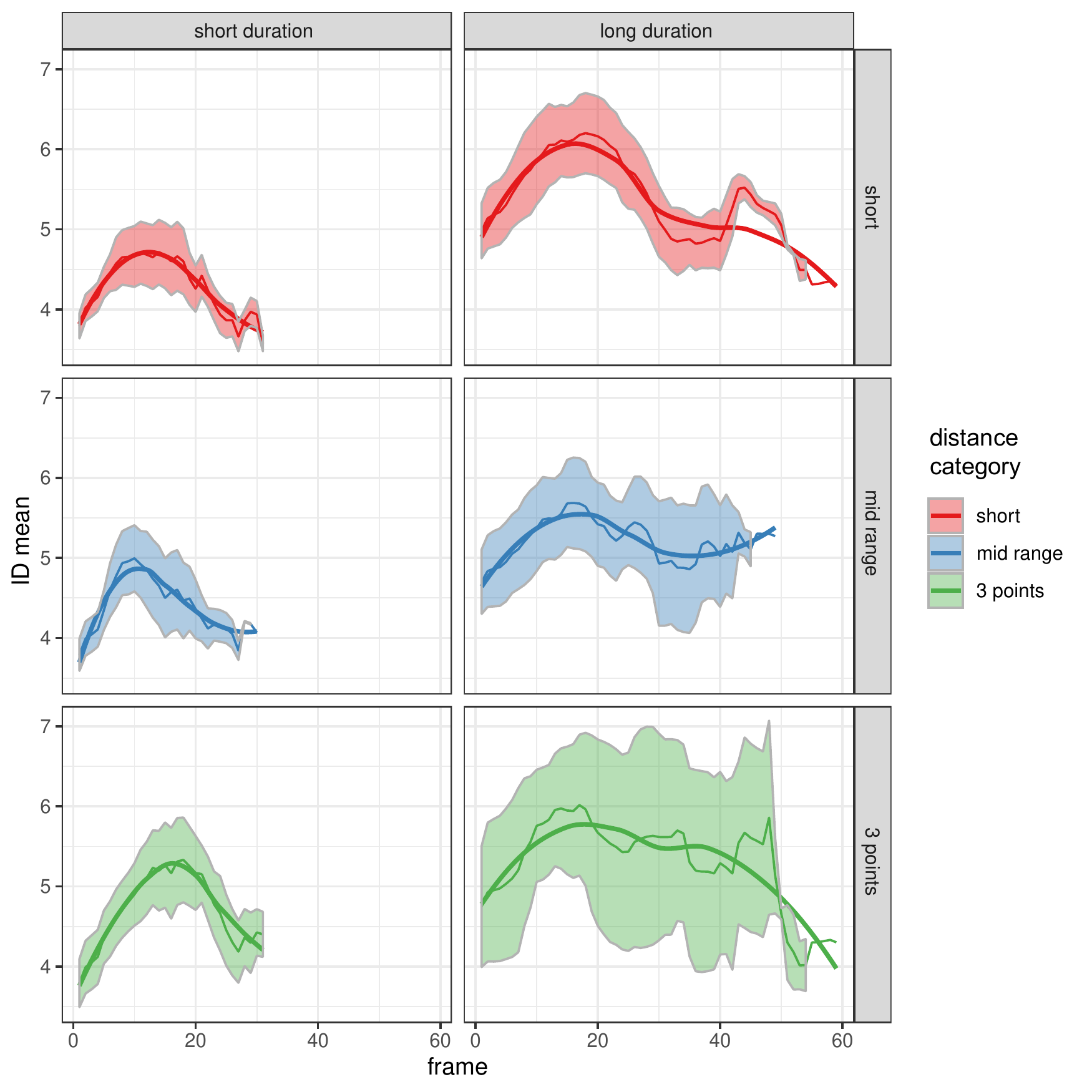}
	\caption{Posterior medians and 95\% credible intervals of the ID during the shots taken by both teams. We consider three distance categories (\texttt{dist\_cat}) short, mid-range and 3-points attempts as well as short (left column) and long duration (right column) possessions ($t \leqslant 12.5$ and $t > 12.5$ seconds respectively).	} 
	\label{fig:type_shot_distance_duration}
\end{figure*}

Additionally, possessions were divided into two groups: short and long duration. We used the cut-off of 12.5 seconds measured from the moment the ball crosses the center line.
Fig.~\ref{fig:type_shot_distance_duration} gives the posterior median of the ID value for the game CLE vs GSW.
The x-axis represents the frame number (2.5frames/second). 
Overall, the ID shows patterns of spikes and declines during the execution of the play. 
%On average, 3 points have a greater intrinsic dimension value compared with short and mid-range shots. 
Short possessions tend to have a peak in ID around frames 10-15 ($\approx$ 4-6 seconds after the ball reaches the offensive court.)
However, for long possession times, the ID reaches the pinnacle at approximately seconds 6-8 or between frames 15-20.

\subsection{Speed and angle}

A potential limitation of the analysis of movement data is that the player locations at time $t$ are not independent of the locations at time $t-1$, $t-2$, etc.
We extended the analysis using the Euclidean distance traveled at every timestamp (speed) by each player and the angle of the trajectory between timestamps. In both cases the original dimension $D = 10$.

Let $y$ and $x$ be the positions in the vertical and horizontal axis on the offensive side of the basketball court respectively. 
The subscript $t$ represents the time stamp of a play.
The speed $s$ (as distance per unit of time) and the angle $\theta$ are then: $s_t = \sqrt{\left(y_{t+1}-y_t\right)^2 + \left(x_{t+1}-x_t\right)^2 }$ and $\theta_t = \tan^{-1}\left(\left(y_{t+1}-y_t\right)/\left(x_{t+1}-x_t\right)\right)$.

\begin{figure}[htp]
\centering
\begin{subfigure}{.5\textwidth}
  \centering
	\caption{Speed} 
	%\vspace{-0.6cm}
		\includegraphics[width=3.5in]{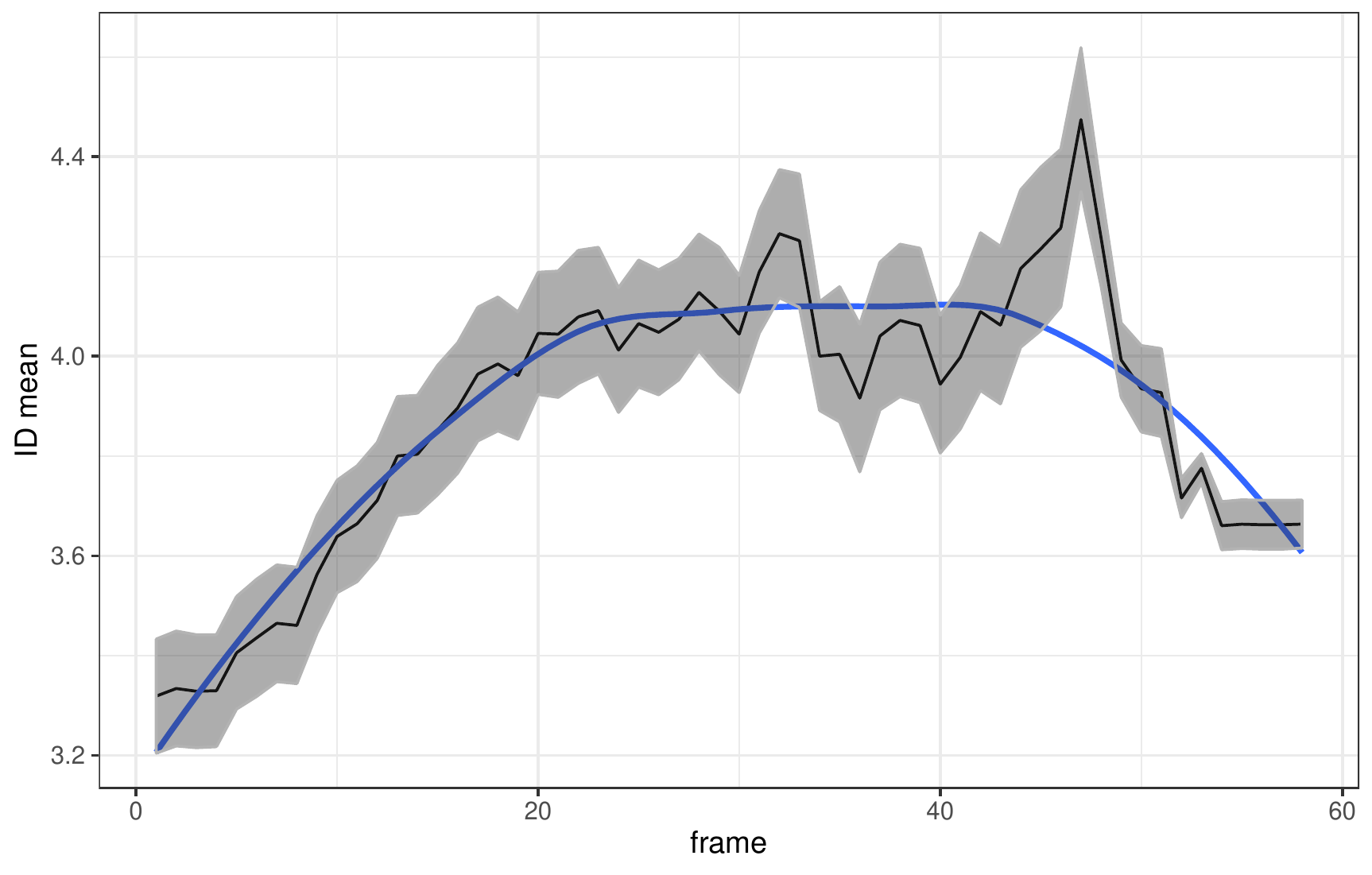} %id25
		\label{fig:dist}
\end{subfigure}%
\begin{subfigure}{.5\textwidth}
  \centering
	\caption{Angle} 
	%\vspace{-0.6cm}
\includegraphics[width=3.5in]{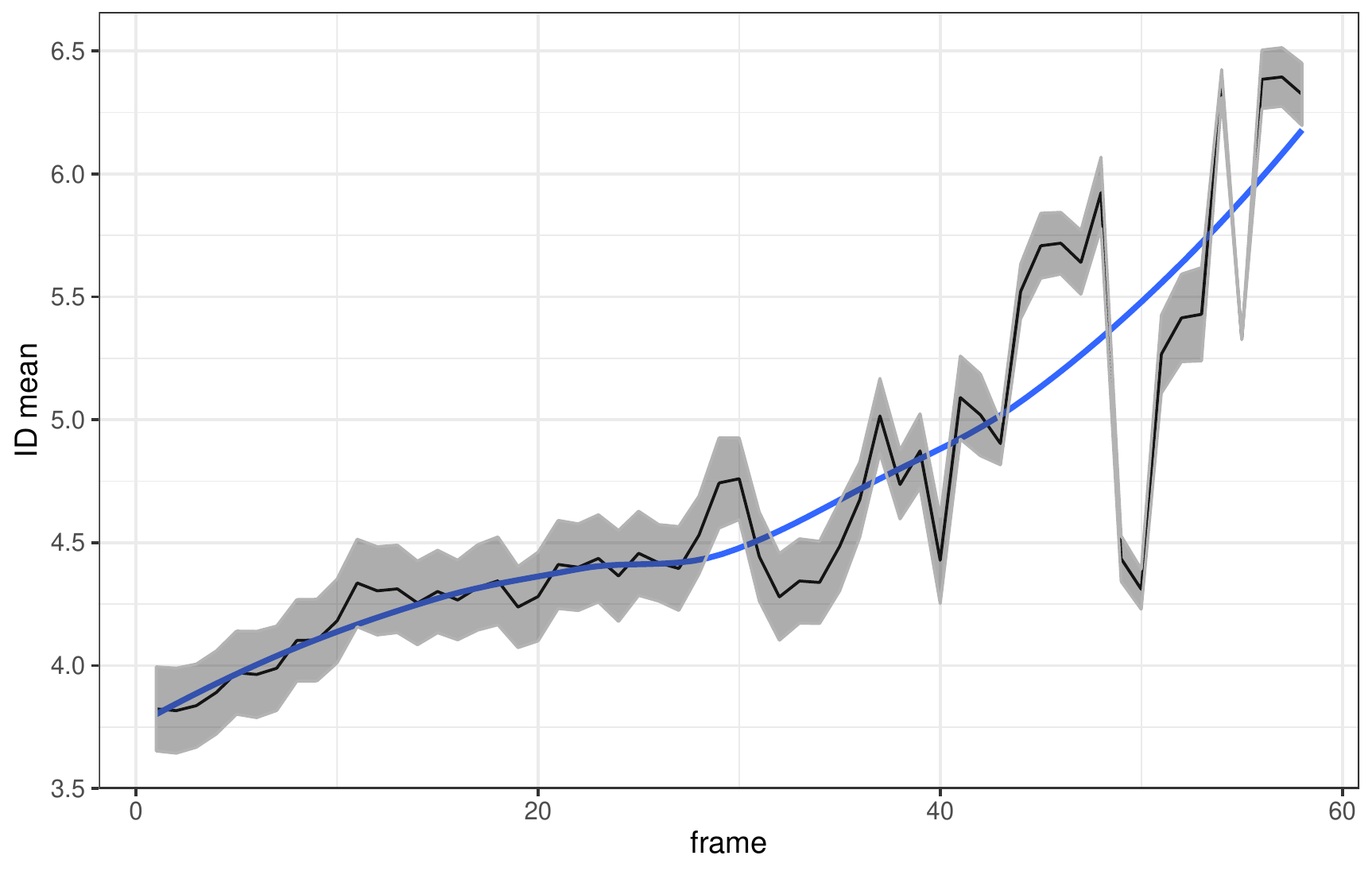} %id145
\label{fig:angle}
\end{subfigure}
\vspace{-0.5cm}
\caption{Posterior medians and 95\% credible intervals of the players speed and angle ID per time frame.}
\label{fig:speed_angle}
\end{figure}

The posterior median of the ID of the speed and angle datasets are shown in Fig. \ref{fig:speed_angle}. We note a gradual increase in ID on the speed from frames 1 to 20 after which it stabilizes. 
The players' directions exhibit different behavior, with a sustained increase during the execution of the play.

\pagebreak

\subsection{ID analysis of shot charts}
\label{sec:charts}

From each play, we inferred the locations of the players at the moment of the shoot.
This point in time was obtained using the $z$ coordinate of the ball (radius). 
We selected the events = $\{$ShotMissed, ShotMade$\}$ and the location of the ball was not considered in the ID estimation.

\subsubsection{Two teams approach}
We compute the intrinsic dimension using the shot chart data from the home and away teams.
We split the data into two sets as follows: (1) field goals shots taken when the home team (e.g. GSW) is attacking and the away team (CLE) is on defense; and (2) field goals shots from the away team (CLE) on attack and the home team on defense (GSW).
The number of rows of each dataset is the number of attempted field shots. 
The number of columns represent the original dimension of the data i.e. $D$ is 20 (2 players' coordinates ($x$ and $y$) $\times$ 5 players $\times$ 2 teams).
The intrinsic dimension for the set of players (5 vs 5) corresponds to the number of independent directions in which the 20-dimensional points are embedded.

\begin{figure*}[ht]
	\centering
		\includegraphics[width=4.85in]{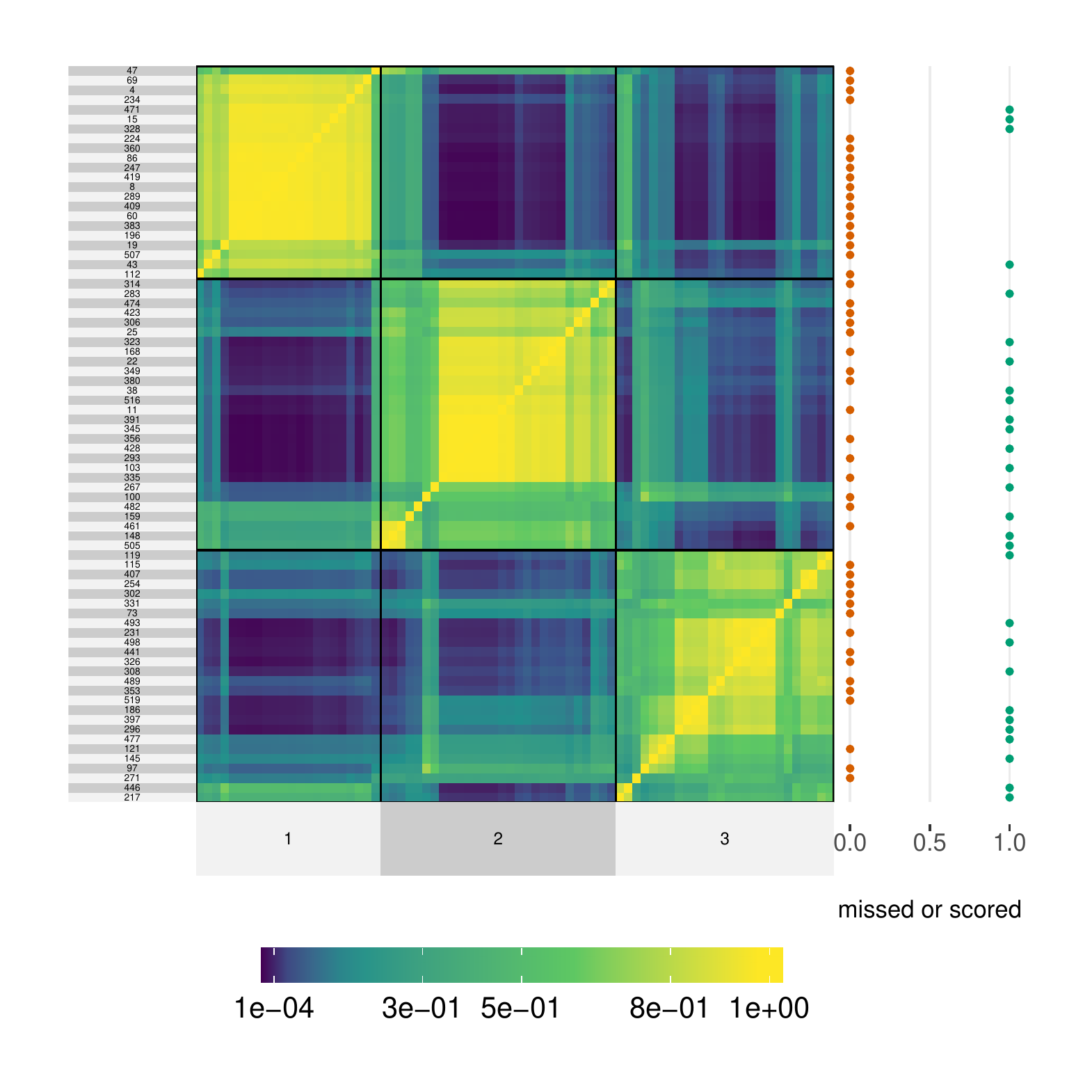}
			\vspace{-0.75cm}	
	\caption{Heatmap of the posterior similarity matrix for the plays where CLE was on attack and GSW on defense. The right dots plot shows the field goals made (green dots) and missed (orange dots).} 
	\label{fig:CLE2teams}
\end{figure*}

In Fig. \ref{fig:CLE2teams} we show a heatmap of the posterior similarity matrix for the plays where CLE was on attack and GSW on defense.
Columns and rows were reordered based on hierarchical clustering so that plays with similar probabilities of belonging to a certain cluster tend to be grouped. 
The labels in the $x$ axis represent the game event. 
Three main clusters are identified (in yellow color representing high probability), with approximately equal number of plays.
For example, $idn = 15$ is the play shown in Fig. \ref{fig:CLE2teams}.
The right hand side dot plot shows the outcome of each of the field goals. 
Missed shots (0) are in orange color and shot that were made (1) are in green. 
We note a large number of unsuccessful plays in cluster 1 (id = 47--112). 

\begin{figure*}[ht]
	\centering
		\includegraphics[width=4.85in]{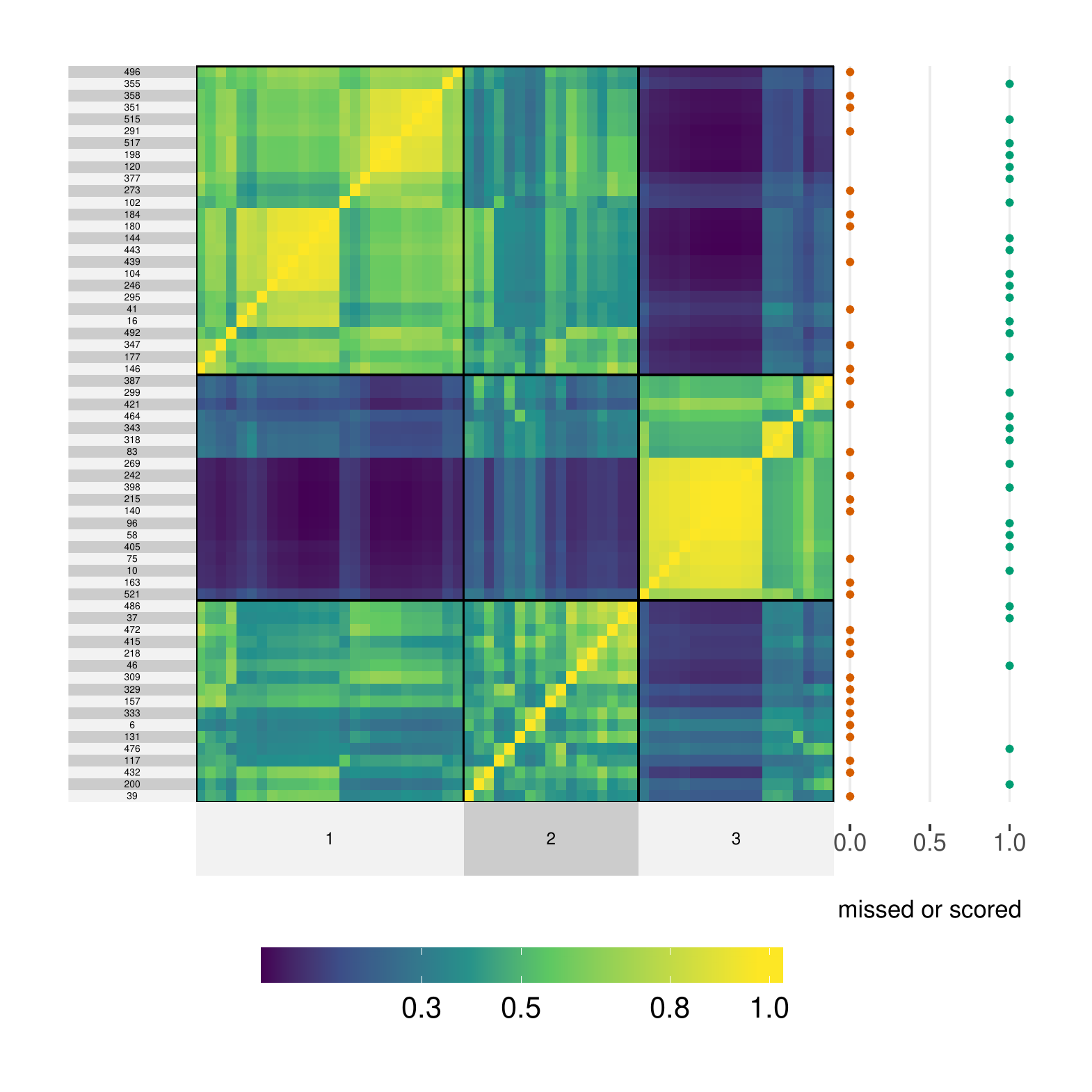}
	\vspace{-0.75cm}	
		\caption{Heatmap of the posterior similarity matrix for the plays where GSW was on attack while CLE was on defense. The right dots plot shows the field goals made (green dots) and missed (orange dots).} %
	\label{fig:GSW2teamsapproach}
\end{figure*}

Equally from the same game we have the field goal plays by GSW, which are represented in Fig.~\ref{fig:GSW2teamsapproach}.
Cluster 2 $(39, 200, \ldots, 486)$ represents a group of plays where GSW had probability of success (22\%), which is well below the other clusters. 

Table~\ref{table:cle} contains the probability of success per shot type (short, mid-range and 3-points) and the median ID value for both teams.

\begin{table}[hptb]
\centering
\caption{Probability of success when each team is on attack per type of shot (short, mid-range and 3-points).
In the season, CLE had a probability of scoring of 0.362 in 3-points and 0.514 in 2 points shots.
GSW's probability of scoring in 3 and 2 points shots was 0.416 and 0.528 respectively. 
}%https://www.basketball-reference.com/leagues/NBA_2016.html
\begin{tabular}{rrrrr}
  \hline

 team & dist\_cat & ns & $p$ success & $\bar{ID}$ \\  
  \hline
CLE & short &   40 & 0.375 & 9.596 \\ 
CLE & mid\_range &   16 & 0.500 & 10.182 \\ 
CLE & 3\_points &   20 & 0.250 & 10.052 \\ \hline
GSW & short &   25 & 0.640 & 11.317 \\ 
GSW & mid\_range &   22 & 0.409 & 11.056 \\ 
GSW & 3\_points &   15 & 0.333 & 11.049 \\ 

  \hline
	\label{table:cle}
\end{tabular}
\end{table}

\begin{comment}

\begin{table}[hpbt]
\centering
\caption{Probability of success when CLE is on attack per cluster. The overall field goal percentage of CLE in the season was 0.460 and the league average was 0.452.}
\begin{tabular}{rrrr}
  \hline
cluster & $n$ & $p$ success & $\bar{ID}$   \\ %& Mean ID
  \hline
1 &   27 & 0.259 & 10.687 \\ 
  2 &   21 & 0.429 & 10.449 \\ 
  3 &   25 & 0.400 & 7.435 \\ 
  4 &    3 & 0.667 & 8.562 \\ 
  \hline
	\label{table:cle}
\end{tabular}
\end{table}

\end{comment}

\begin{comment}

\begin{table}[hpt]
\centering
\caption{Probability of success when GSW is on attack. The probability of scoring of GSW in the season was 0.487 and the league average was 0.452.} %https://www.basketball-reference.com/leagues/NBA_2016.html
\begin{tabular}{rrrr}
  \hline
cluster & $n$ & $p$ success & $\bar{ID}$   \\ %& Mean ID
  \hline

  1 &    6 & 0.333 & 11.837 \\ 
  2 &   13 & 0.462 & 8.197 \\ 
  3 &   37 & 0.486 & 12.481 \\ 
  4 &    6 & 0.667 & 9.713 \\
	
   \hline
	\label{table:gsw}
\end{tabular}
\end{table}

\end{comment}

\begin{figure*}[h]
	\centering
		\includegraphics[width=4.25in]{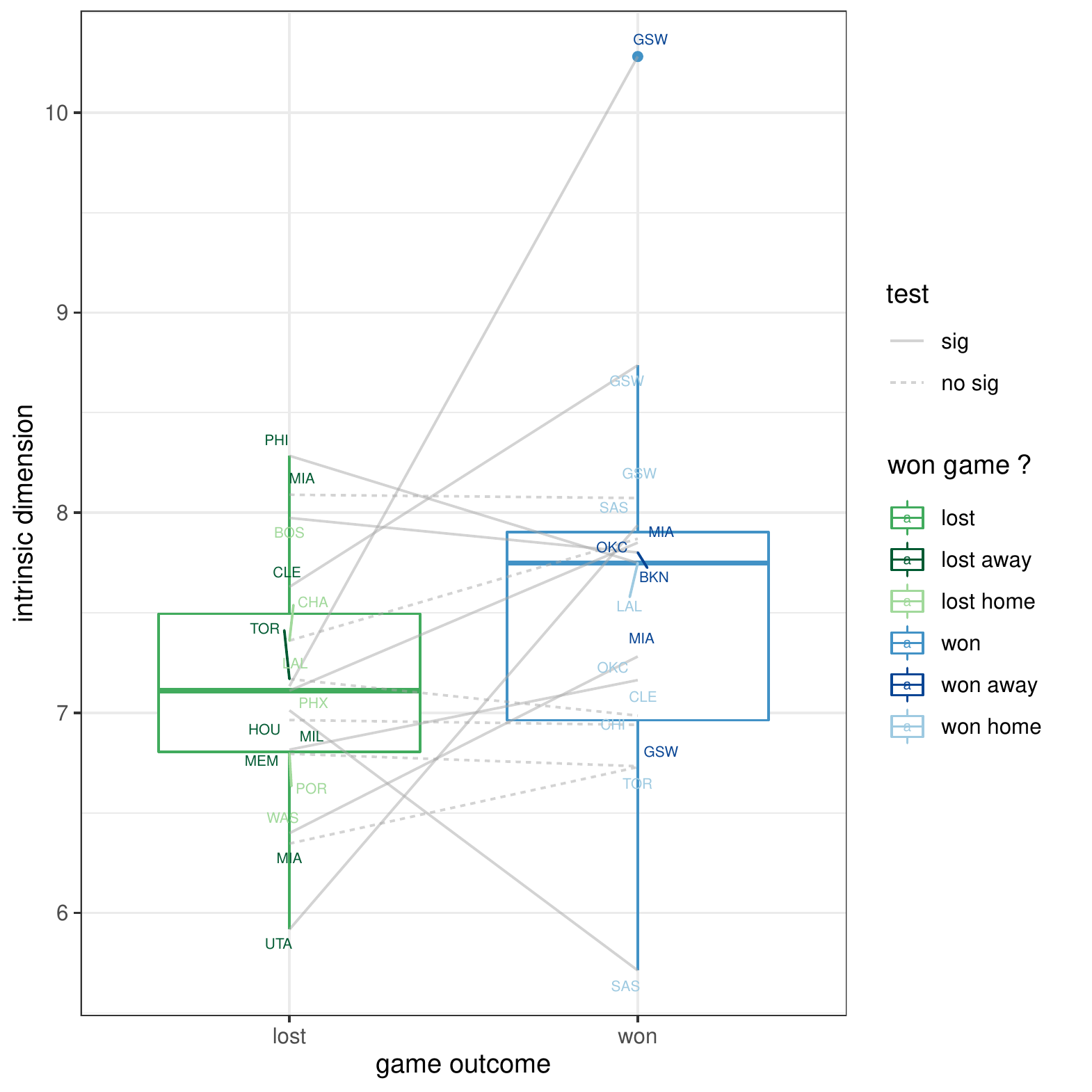}
	\caption{Boxplots of the posterior intrinsic dimension of the winners and losers in the 15 matches. Each game is represented by a gray line connecting the two involved teams. Solid line between two teams represents significant difference between the posterior median ID using a Mann-Whitney test } 
	\label{fig:id15teams}
\end{figure*}

We computed the ID values for the shots taken during 15 games of the season. Although the number of games is relatively small, there seems to be a positive association between the overall posterior median of the intrinsic dimension and the game outcome. 
The boxplots in Fig. \ref{fig:id15teams} show the ID for winning and losing teams. Each game is represented by a gray line connecting both teams from the left to the right boxplot.
The solid line signals the games where a significant difference was found between the posterior medians using a Mann-Whitney test. 
The dashed line represents no evidence showing differences between the teams.
In six of these games, the winner had a greater intrinsic dimension. 
In six, there was no difference in the ID between winners and losers and in three cases the losers had higher ID values.
 
\begin{figure*}[hptb]
	\centering
		\includegraphics[width=4in]{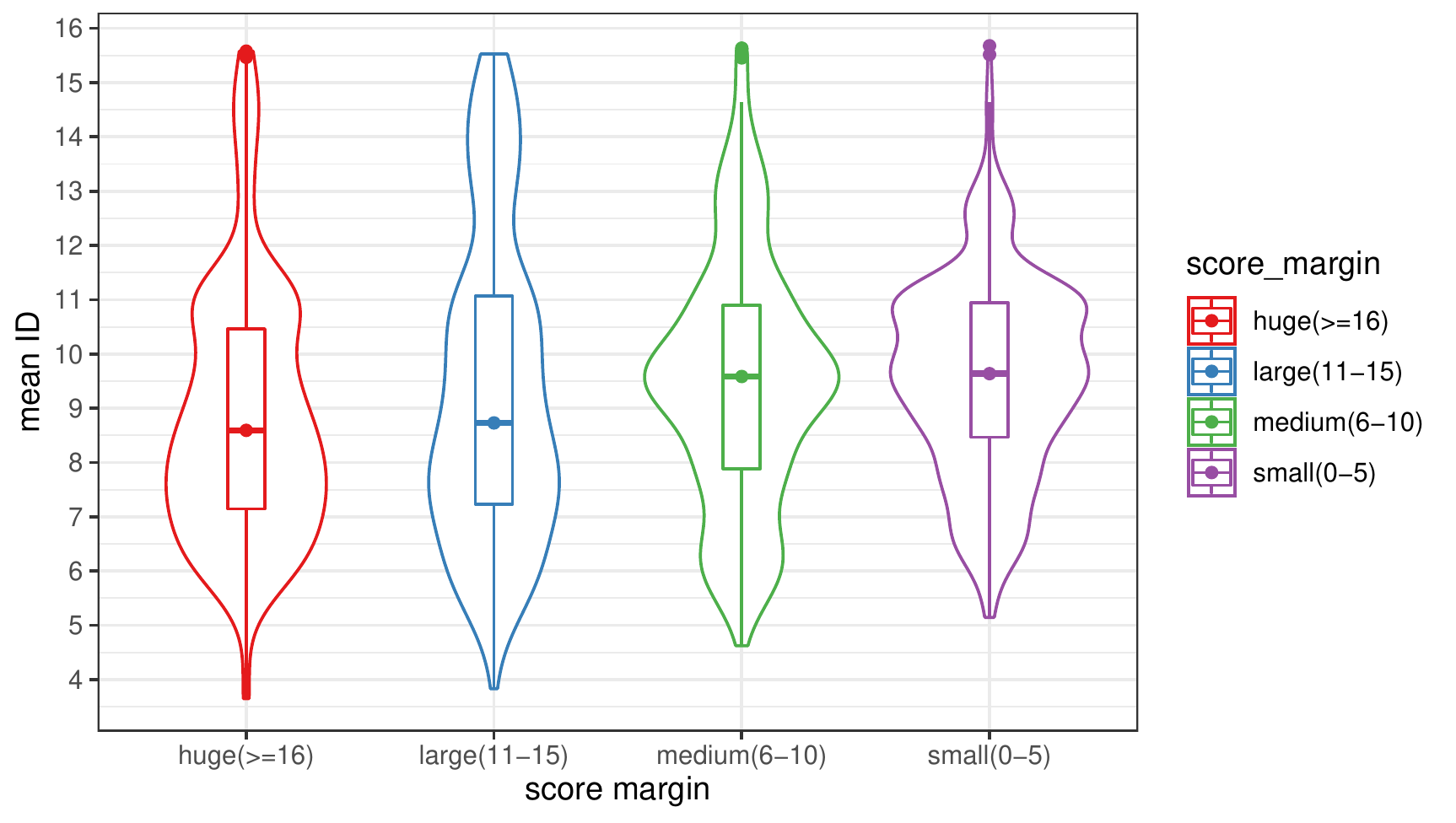}
	\caption{Violin/boxplot of the posterior intrinsic dimension as a function of the score margin in the 15 matches.  } 
	\label{fig:score_margin}
\end{figure*}

 % .
Plays generally follow the path of less resistance.
We argue that plays tend to have an increased movement complexity when the difference in the score is small, usually as a result of a tighter defense. 
In Fig. \ref{fig:score_margin} we show the distributions of the ID's posterior medians for different score margins categories \{small(0-5 points), medium(6-10 points), large(11-15 points) and huge($>=$16 points)\}.
Pairwise comparisons using the Wilcoxon rank sum test shows evidence supporting the argument that the smaller score margin the greater the ID and the complexity.

\begin{table}[hpt]
\centering
\caption{Pairwise comparisons (p-values) of the distributions of the ID for different scoring margins based on the Wilcoxon rank sum test. The alternative hypothesis is: the category in the rows has greater median ranks than the one in the columns.} %produced file 81 utils.R

\begin{tabular}{rrrr}
  \hline
 & huge($>$=16) & large(11-15) & medium(6-10) \\ 
  \hline
large(11-15)   & 0.285 &  &  \\ 
  medium(6-10) & $<$0.0001 & 0.334 &  \\ 
  small(0-5)   & $<$0.0001 & 0.010 & 0.505 \\ 
   \hline
	
   \hline
	\label{table:pval}
\end{tabular}
\end{table}

%\clearpage

\subsection{Individual Team Approach}

We ran a similar analysis for individual datasets comprising the locations of the 5 players from each team in attack and then when they are in a defensive role. %Similarly, the computation is done for the visitor team.
In this case the dimension is $D = 10$ (location in $x$ and $y$ $\times$  5 players).
This analysis yields for each team clusters of shot charts plays with a low and a high return in offense and defense.

\begin{figure}[h]
\centering
\begin{subfigure}[b]{0.55\textwidth}
  \centering
	\caption{Attack} 
		\includegraphics[width=3.5in]{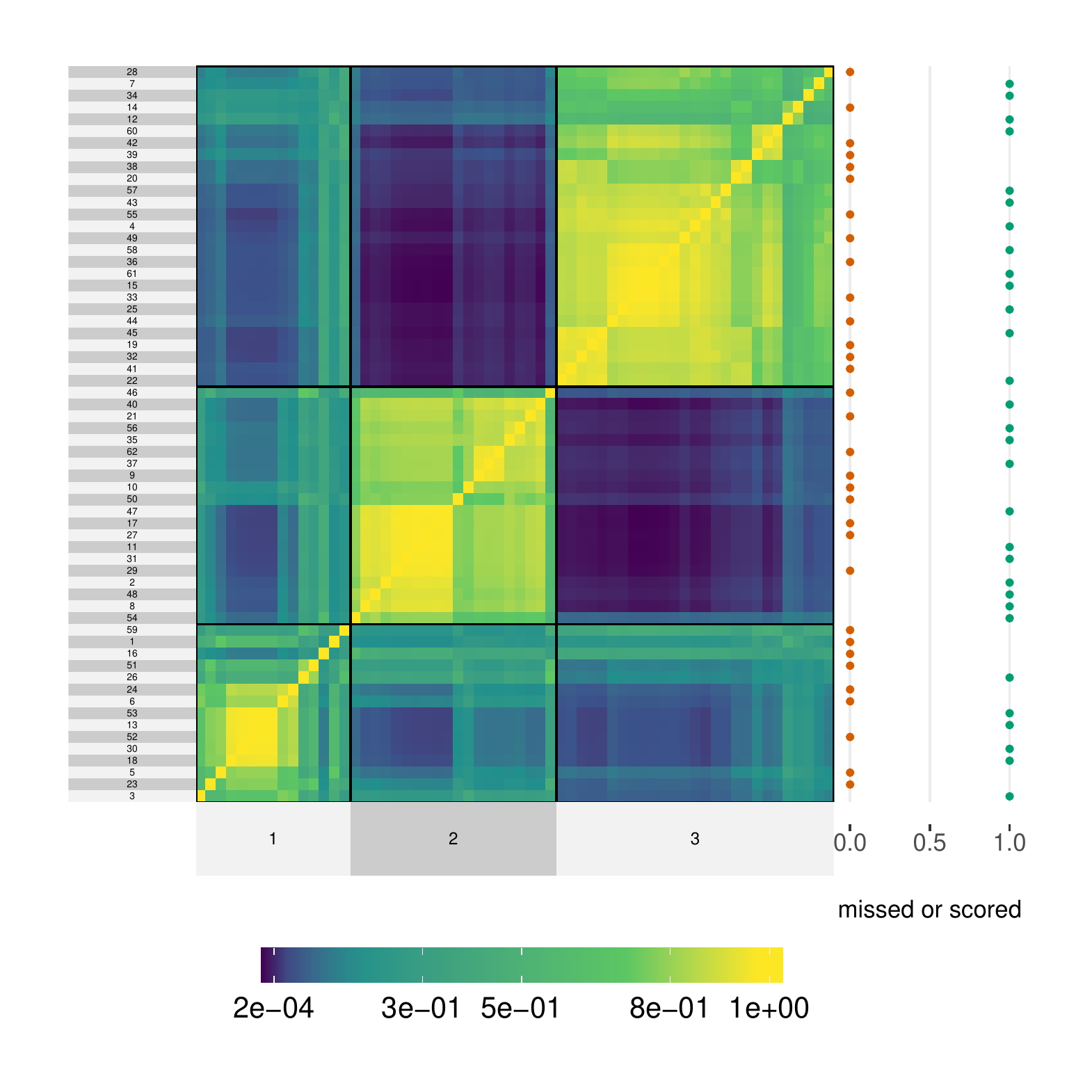}
  \label{fig:HOME_attacking.pdf}
\end{subfigure}%
\begin{subfigure}[b]{0.5\textwidth}
  \centering
	\caption{Defense} 
		\includegraphics[width=3.5in]{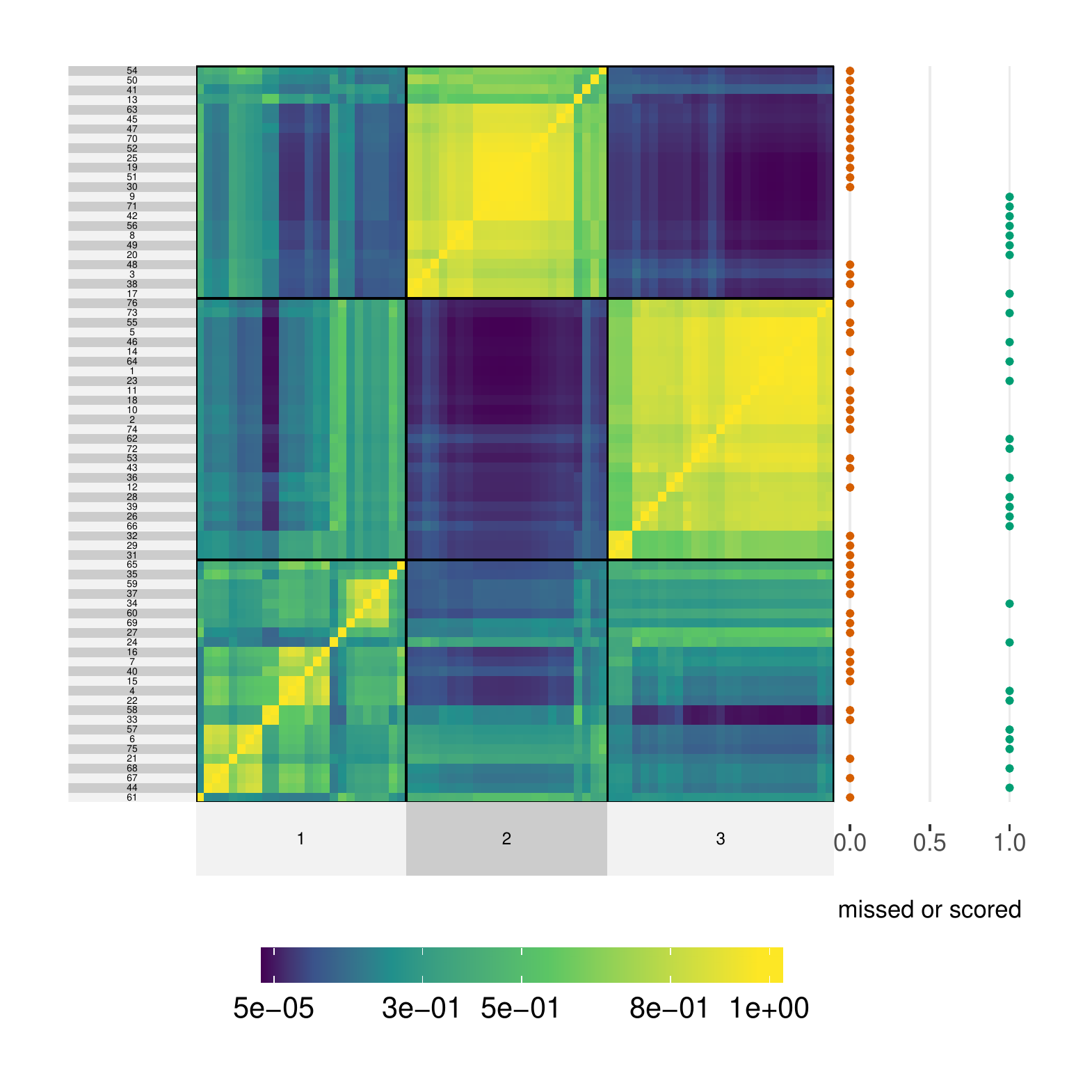}
  \label{fig:HOME_deffending.pdf}
\end{subfigure}
\caption{Heatmap and clusters of the shot chart plays by GSW in attack (a) and defense (b). 
The x-axis gives the cluster and the y-axis represent the play number.
The top dots plot shows the field goals made (green dots) and missed (orange dots).}
\label{fig:indivGSW}
\end{figure}

Fig~\ref{fig:indivGSW} shows the posterior similarity heatmaps of the plays by GSW.  
On each of the plots, we defined three clusters. 
For instance, in subfigure (a) finding cluster 1
 in the x-axis, we find that plays  $59, 1, $\ldots$, 23$ and 3 have a large probability of belonging to this cluster (in yellow color).
The outcome of each play is represented in the dot plot on the right-hand side.  
Table~\ref{table:propo} gives the proportion of successful plays in the clusters.  
56\% of these offensive shots in cluster 1 (a) were successful.
Similarly, only 16.7\%  of the attacking plays in the second cluster were scored.   

In (b), we show the clusters from the defensive placements of GSW. For example, cluster 3 shows a poor defensive outcome for GSW, allowing 55\% scoring by CLE. 

\begin{table}[h]
\centering
\caption{Probability of success in the offensive and defensive roles for GSW.}
\begin{tabular}{rrr}
  \hline
role & cluster & $p$ success \\ 
  \hline
attack & 1 &     0.400 \\ 
attack & 2 &     0.550 \\ 
attack & 3 &     0.481 \\ \hline
%attack & 4 &     0.500 \\ \hline
 defense &  1 &     0.360 \\ 
  defense &  2 &   0.333 \\ 
  defense &  3 &     0.407 \\ 
  %defense &  4 &     0.375 \\ 
   \hline
	\label{table:propo}
\end{tabular}
\end{table}

\begin{figure}[htp]
\centering
\begin{subfigure}[b]{0.55\textwidth}
  \centering
	\caption{Attack} 
		\includegraphics[width=3.5in]{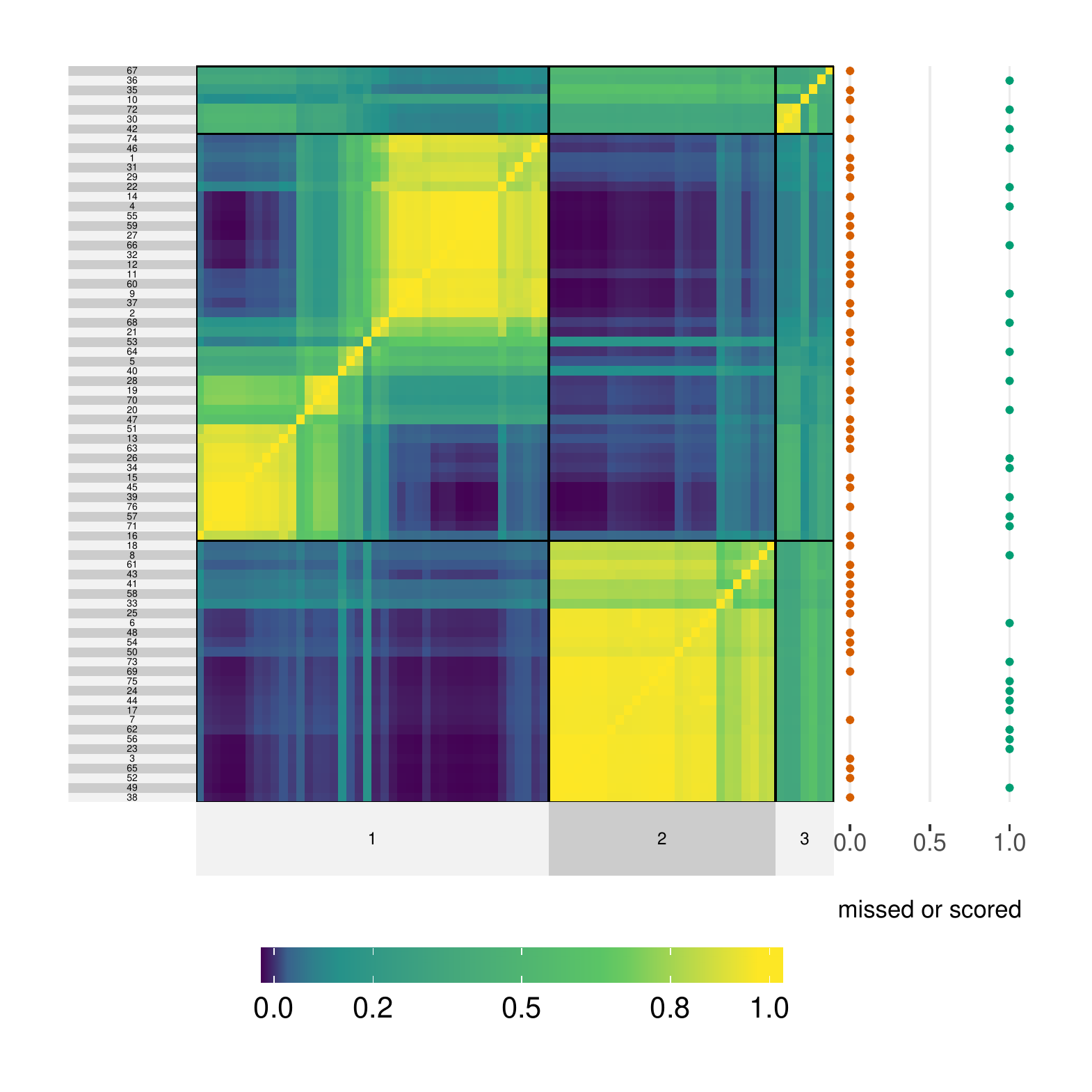}
  \label{fig:AWAY_attack.pdf}
\end{subfigure}%
\begin{subfigure}[b]{0.5\textwidth}
  \centering
	\caption{Defense} 
		\includegraphics[width=3.5in]{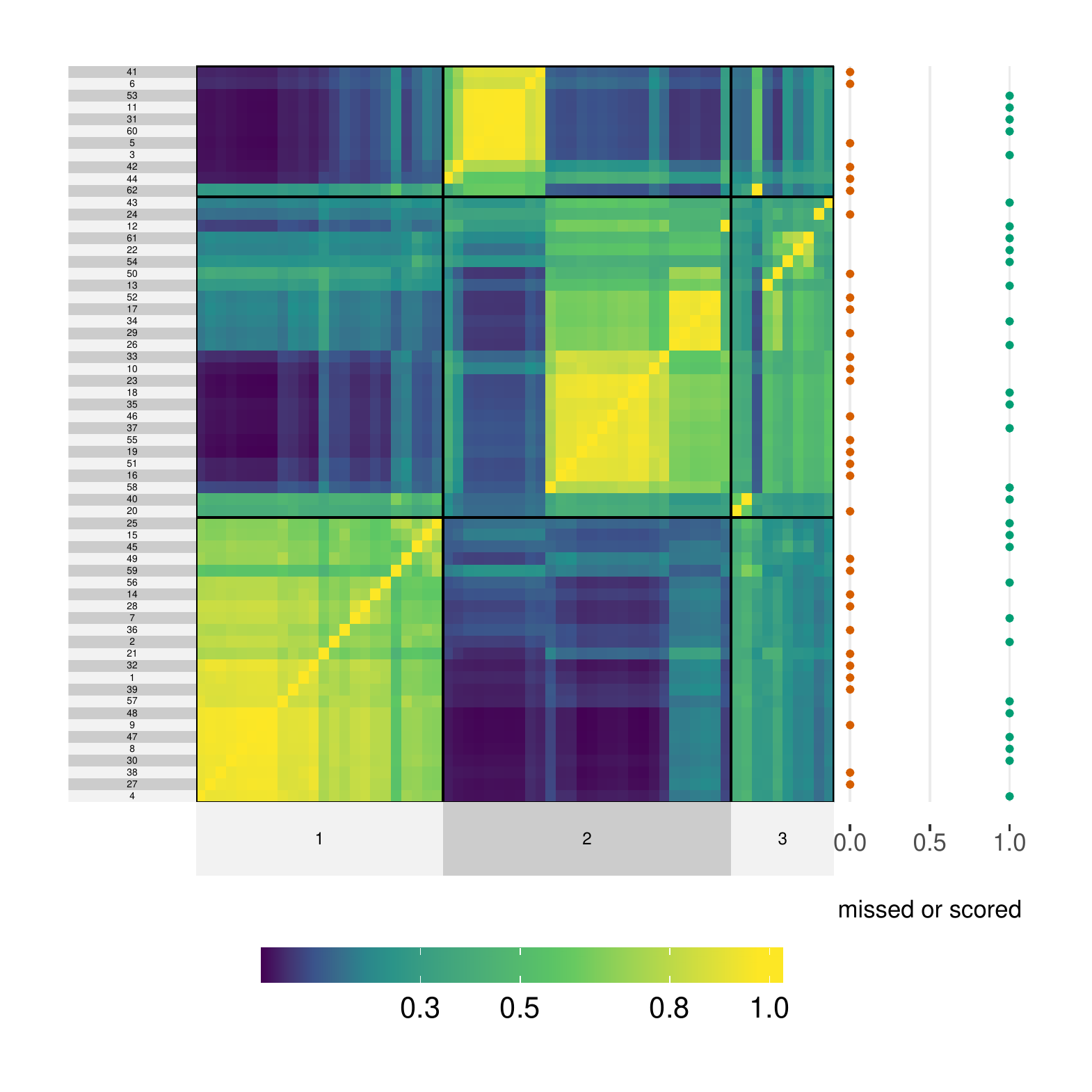}
  \label{fig:AWAY_deffence.pdf}
\end{subfigure}
\caption{Heatmap and clusters of the shot chart plays by CLE in attack and defense. 
The x-axis gives the cluster and the y-axis represent the play number.
The top dots plot shows the field goals made (green dots) and missed (orange dots).}
\label{fig:indivCLE}
\end{figure}

Furthermore in Fig~\ref{fig:indivCLE} we present the posterior similarity of CLE in attack (a) and defense (b).
From (b) cluster 3 contains six plays where the defense by CLE was ineffective allowing 83\% success for GSW.

\begin{table}[hpt]
\centering
\caption{Probability of success in the offensive and defensive roles for CLE.}
\begin{tabular}{rrr}
  \hline
role & cluster & $p$ success  \\ 
  \hline
attack & 1 &    0.333 \\ 
attack &  2 &    0.407 \\ 
attack &  3 &    0.429 \\ \hline
%attack &  4 &     0.250 \\ \hline

 defense &  1 &     0.500 \\ 
  defense &2 &     0.429 \\ 
 defense & 3 &      0.600   \\ 
 % defense &4 &     0.414 \\ 
   \hline
	\label{table:propo}
\end{tabular}
\end{table}

\section{Discussion and conclusions}
\label{dc}

The advent of sports tracking technology is flooding sports analytics and sports science with large datasets \citep{lazar2014big}.
Especially in basketball, massive datasets are generated on each game,  making increasingly challenging and laborious for individual analysis and for making meaningful inferences of players/teams performance and for obtaining competitive advantages.

As a result, researchers and practitioners are resorting to multivariate statistical analysis so that high dimensional data can be reduced, handled and interpreted more conveniently.    
The purpose of the current study was to present a different perspective in the analysis of high-resolution player tracking data from the NBA.

We used the intrinsic dimension of the player's positions $\left(x,y\right)$ in Cartesian coordinates to: 
\begin{itemize}
    \item determine different stages in the execution of offensive actions. %dynamic  
    \item identify clusters in shot chart data.
    \item compare and assess the relationship between intrinsic dimension and game performance.
 \end{itemize}

We employed a local model-based approach for ID estimation developed by \citet{allegra2019clustering} because it has been found to be fast and accurate. We have proposed different enhancements, ranging from the choice of more meaningful prior distributions to a better postprocessing of the MCMC output. 
The results show that using this Bayesian clustering approach we can satisfactorily identify plays with a lower and higher than average return.
Our method could help coaches to plan more effective attacking and defensive plays. 

Higher games' median ID values were found to be linked to a higher play uncertainty in the attack. This results are in line with previous findings, see e.g. \citet{hobbs2018playing} that discusses ball entropy.  
We also found that a larger ID in shot charts is positively associated with winning games. This claim needs to be validated using a larger sample size though. 

This approach also enhances our understanding of how players' moving tactics impact the outcome of a play. 
Stages like ball handling, creating space for passing, shooting and following through have different characteristics and 
can be identified using the coclustering matrix along with the median ID curve. %Another important finding was that s
An increase in ID values was found when the players are creating an opportunity for passing and shooting. This is expected as players on attack tend to move with larger uncertainty and entropy using for example screen actions.
Similarly, plays show a decline in ID near the end when the players are following through shots or returning to the opposite part of the court.

This Bayesian approach could complement manual video game analysis, providing effective and fast clustering.
In addition, these analyzes can be easily extended to other sports like football and rugby that have implemented player tracking technology. 
 However, we are well aware of some of the limitations of this approach. First, the choice of $K$, the number of mixture components, does not take into account any form of uncertainty. We are working on a Bayesian nonparametric extensions to solve this issue, such as Dirichlet Process Mixture models \citep{Antoniak1974,Escobar2012}. Moreover, the analysis of how the ID changes across time frames provides remarkable results, but does not satisfy the hypothesis of independence across the observation. This issue paves the way to an interesting research path, where the model-based ID estimation framework can be combined with Hidden Markov Models \citep{Baum1966}. 
Further research should be undertaken to assess the link between intrinsic dimension and issues like player energy consumption \& fatigue, movement dynamics.

\section*{Acknowledgement}
This research was supported by the Australian Research Council (ARC) Laureate Fellowship Program, the Centre of Excellence for Mathematical and Statistical Frontiers (ACEMS) and by the project ``Bayesian Learning for Decision Making in the Big Data Era'' (ID: FL150100150). First Investigator: Prof Kerrie Mengersen.
We thank Wade Hobbs for his comments and suggestions.
All computations and visualizations were carried using \textsf{R} using the packages \textsf{mcclust} \citep{mcclust}, \textsf{superheat} \citep{superheat}, \textsf{tidyverse} \citep{tidyverse} and 
\textsf{gganimate}
\citep{gganimate}.

%\pagebreak 

%\Appendix
\appendix

%\begin{comment}
%\end{comment}  %----------------------------------------

\section{Posterior ID values based on three priors}
\begin{figure}[ht]
    \centering
    \includegraphics[scale=.75]{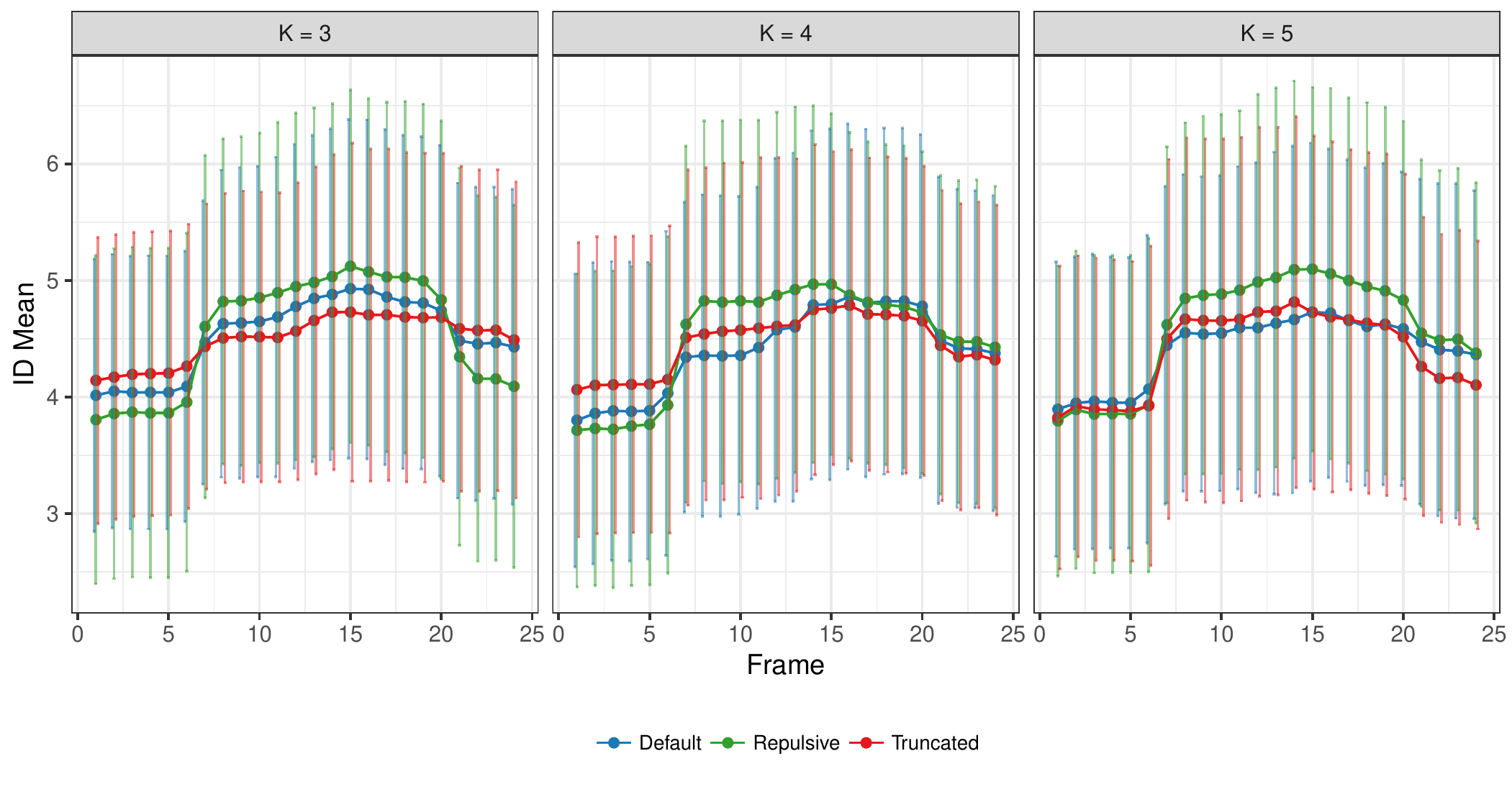}
    \caption{Comparison of the posterior ID values for the three methods and several mixture components.}
    \label{fig:K_method_ID}
\end{figure}

\section{Example of the ID in event 217 of the game (Irving's driving bank shot. Score = 39 vs 37).} 
\label{sec:ap21}

\begin{figure}[hbp]
\centering
\begin{subfigure}[b]{0.5\textwidth}
  \centering
	\caption{Movement of three players and the ball.} 
		\includegraphics[width=4.4in]{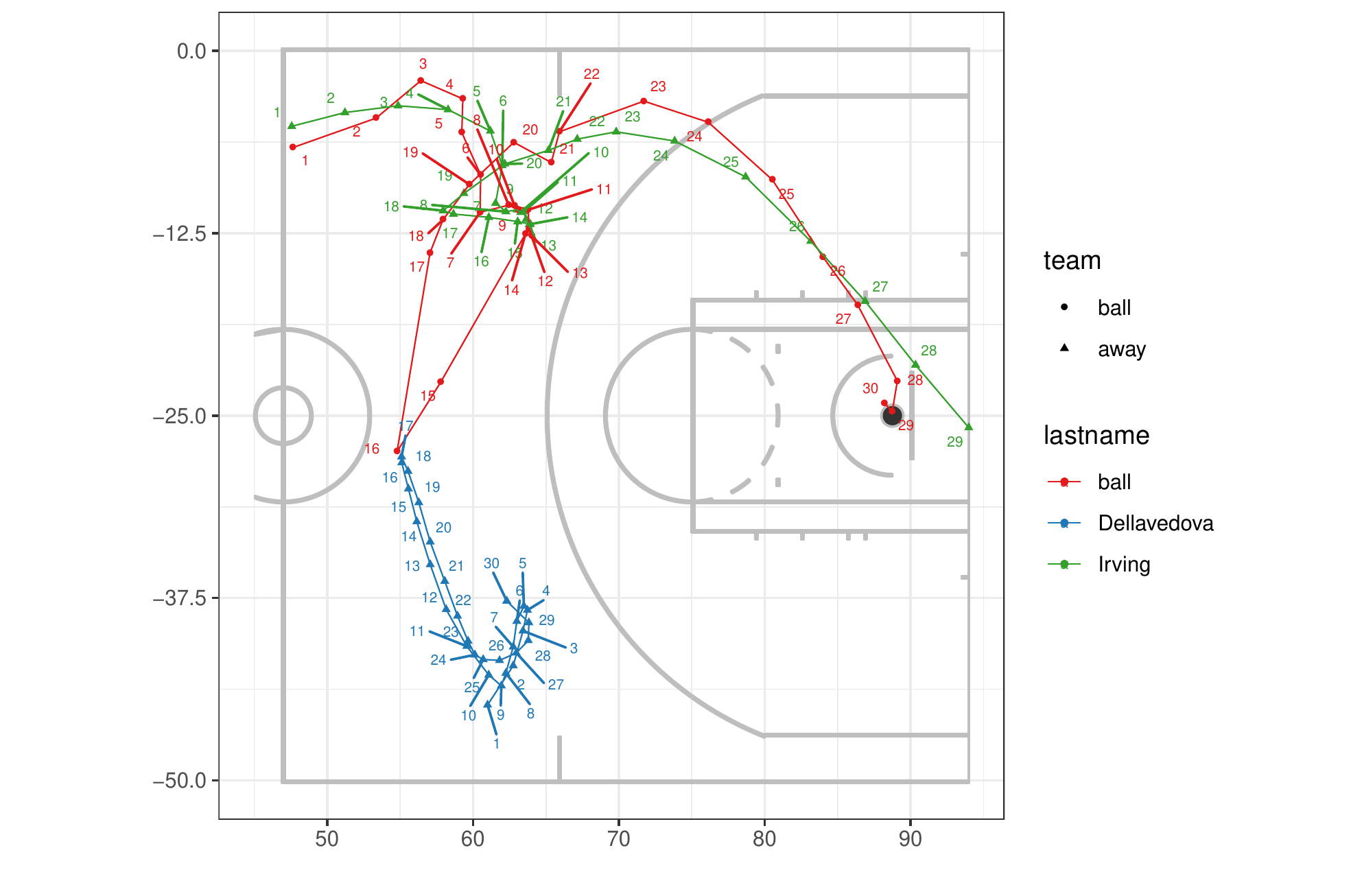}
  \label{fig:id217traj}
\end{subfigure}%
\begin{subfigure}[b]{0.5\textwidth}
  \centering
	\caption{Heatmap of the 30 time stamps during the play and posterior medians of the ID.} 
	\vspace{-0.85cm}
		\includegraphics[width=3.45in]{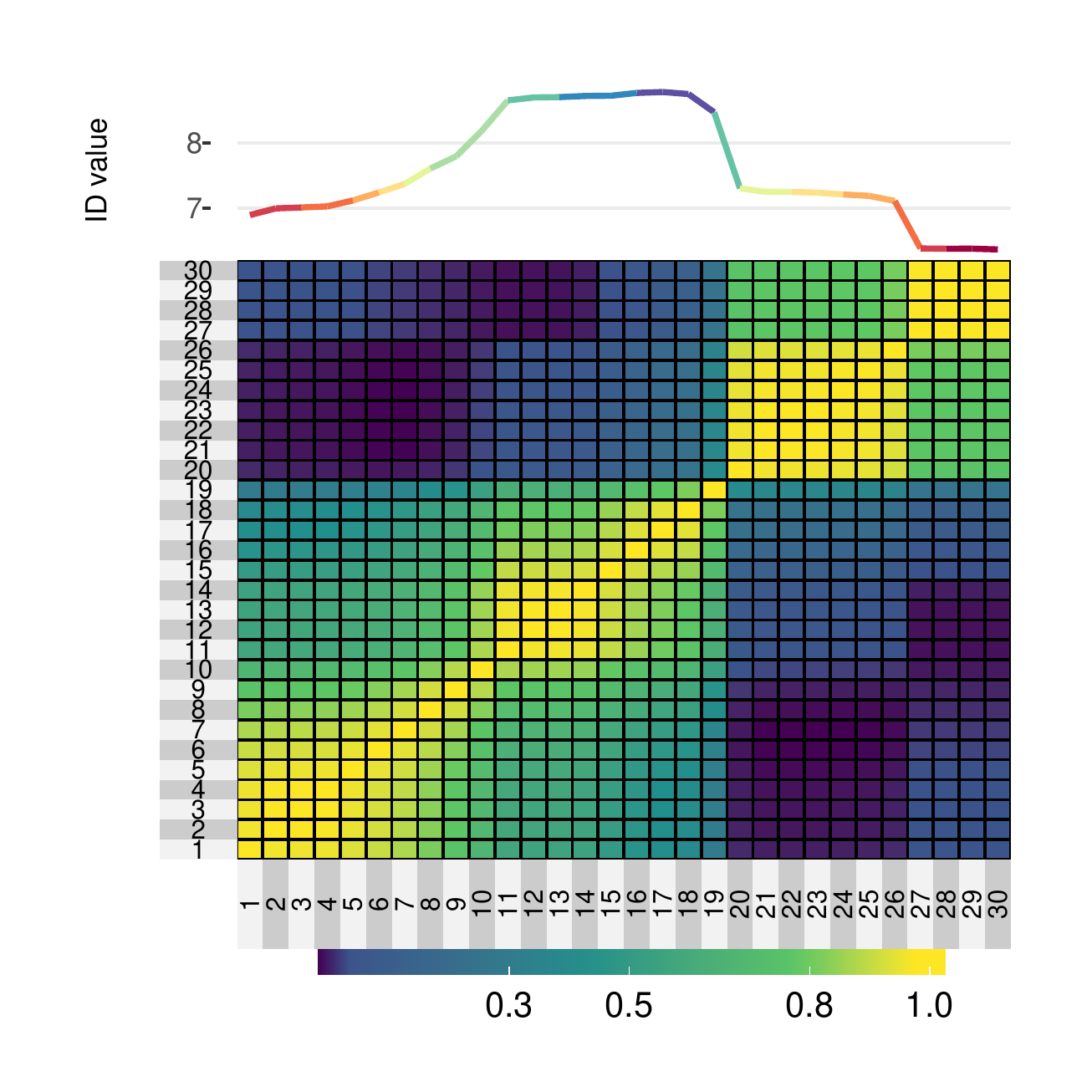}
  \label{fig:id217superheat}
\end{subfigure}
\caption{Trajectory of the players/ball and ID in a two-point driving bank shot. \url{https://youtu.be/jb57MFQLoRo?t=180}}
\label{fig:both}
\end{figure}

\clearpage

%% if your bibliography is in bibtex format, uncomment commands:
\bibliographystyle{imsart-nameyear} 
% Style BST file (imsart-number.bst or imsart-nameyear.bst)
%\bibliography{bibliography}       % Bibliography file (usually '*.bib')

\bibliography{ref}

%% or include bibliography directly:
% \begin{thebibliography}
% \bibitem[\protect\citeauthoryear{???}{???}]{b1}
% \end{thebibliography}

\end{document}